\pdfoutput=1
\documentclass[aps, prd, twocolumn, nofootinbib, floatfix]{revtex4-1}

\usepackage{epsfig}
\usepackage{hyperref}
\usepackage{amssymb}
\usepackage{color}
\newcommand{\be}{\begin{equation}}
\newcommand{\ee}{\end{equation}}
\newcommand{\ba}{\begin{eqnarray}}
\newcommand{\ea}{\end{eqnarray}}
\newcommand{\bd}{\begin{displaymath}}
\newcommand{\ed}{\end{displaymath}}

\newcommand{\commentout}[1]{{}}

\def\thalf{{\textstyle{\frac{1}{2}}}}

\def\rootg{\sqrt{-g}}

\def\DL{\mathcal{D}_L}

\def\pbulk{\rho^{\rm bulk}}
\def\pshear{\rho^{\rm shear}}
\def\Z{\mathcal{Z}}
\def\Q{\mathcal{Q}}
\def\E0{\mathcal{E}^{(0)}}
\def\e1{\mathcal{E}^{(1)}}

\begin{document}
  \title{{\bf Bulk spectral functions in single and multi-scalar gravity duals}}
  
  \author{Todd Springer}
  \email{springer@physics.mcgill.ca}

  \author{Charles Gale}
  \email{gale@physics.mcgill.ca}

  \author{Sangyong Jeon}
  \email{jeon@physics.mcgill.ca}

  \affiliation{Department of Physics, McGill University, Montreal, Quebec, H3A2T8, Canada}
  \date{December 22, 2010}

  
  \begin{abstract}
    We examine two point correlation functions involving the trace of
    the energy momentum tensor in five-dimensional dual gravity 
    theories supported by one or more scalar fields.  A prescription
    for determining bulk channel spectral functions is developed.
    This prescription generalizes previous work which centered on one
    scalar field.  As an application of these techniques, we
    investigate the bulk spectral function and corresponding sum rule
    in the Chamblin-Reall background.  We show that, when expressed in
    terms of the beta function, the sum rule for the Chamblin-Reall
    background can be written in a form similar to the sum rule in
    Yang-Mills theory.
  \end{abstract}
\maketitle

\section{Introduction}
In recent years, there has been great interest in the correspondence
between certain strongly coupled gauge theories and extra-dimensional
gravity theories.  This duality, originally inspired by the AdS/CFT
(Anti de-Sitter/Conformal Field Theory) correspondence
\cite{Maldacena:1997re,Witten:1998qj,Gubser:1998bc,Klebanov:1999tb}
has led to a set of tools which allow for the computation of physical
observables in some strongly coupled theories.  These tools have
renewed interest in the possibility of a nonperturbative description
of QCD and the strongly coupled matter created in heavy ion
collisions, the QGP (quark-gluon plasma).  Many excellent reviews on
the vast literature of this subject are now available, among them are
\cite{Aharony:1999ti,Son:2007vk,Erdmenger:2007cm,Myers:2008fv,Gubser:2009md}.

In order to describe theories which appear in the real world, such as
QCD, one must move beyond scale invariant theories to nonconformal
ones.  There are several known examples of string theory setups which
give rise to a dual nonconformal field theory
\cite{Klebanov:2000nc,Klebanov:2000hb,Pilch:2000ue,Mia:2009wj}.  An
alternative, phenomenologically based approach involves constructing
an effective five-dimensional gravity background and \emph{assuming}
that it has a dual gauge theory description.  It is also assumed that
all of the AdS/CFT machinery still works for this phenomenological
setup.  The simplest of such backgrounds involve Einstein gravity, and
one or more interacting scalar fields.  Such models have been 
studied extensively in the literature \cite{Gubser:2008sz,
  Gubser:2008ny, Gursoy:2007cb, Gursoy:2007er, Gursoy:2008za,
  Gursoy:2009kk, Springer:2008js,
  Springer:2009wj,Batell:2008zm,dePaula:2008fp,Alanen:2009xs}.  To
date, most phenomenological models have included a single scalar field
for simplicity.  However, it seems necessary to include two scalar
fields to properly describe both conformal and chiral symmetry
breaking \cite{Erlich:2005qh,Batell:2008zm,Kapusta:2010mf}.
Furthermore, many string theoretical setups can be reduced to
effective five-dimensional backgrounds supported by multiple scalar
fields \cite{Benincasa:2005iv, Aharony:2005zr}

In this work, we will use such multi-scalar gravity theories to study
two point functions of the energy-momentum tensor in the dual thermal
gauge theory.  Such correlation functions are interesting because
they allow for the computation of transport coefficients (like shear
and bulk viscosities).  Transport coefficients can be read off from
the low energy behavior of the spectral density (the imaginary part of
these correlation functions).  These transport coefficients are useful
in hydrodynamic descriptions of the QGP, and are not otherwise
calculable perturbatively since the QGP is strongly coupled near the
phase transition.  In addition to AdS/CFT, there have been attempts to
extract transport coefficients from the lattice
\cite{Meyer:2007ic,Meyer:2010ii, Kharzeev:2007wb}. Typically one needs
to include an ansatz for the spectral density in
order to perform such an extraction.\footnote{See \cite{Moore:2008ws}
  for a critical examination of the methodology of such an
  extraction.}  In this case, insight gained from gauge/gravity
duality into the behavior of spectral functions at strong coupling
could be useful.

A prescription for determining the bulk spectral density in a
single-scalar gravity dual was given in \cite{Gubser:2008sz}.  One of
the purposes of the current work is to generalize this prescription to
a matter sector which includes \emph{multiple} scalar fields.  The
results of \cite{Gubser:2008sz} employed a coordinate system where the
scalar field was identified with the radial coordinate.  While this
simplified the analysis, it made the generalization to multiple scalar
fields difficult.  In the present work, we will choose our coordinates
and gauges differently so that generalization to multiple scalar
fields is straightforward.  It should be noted that a different method
for determining the bulk viscosity in multi-scalar gravity dual
theories was developed recently in \cite{Yarom:2009mw}, however our
method allows for the calculation of the spectral function at all values
of the frequency $w$.

As an example application, we will determine the bulk spectral density
in a particular single scalar setup called the Chamblin-Reall
background \cite{Chamblin:1999ya}.  This background is one of the only
known nonconformal setups where exact results for the bulk viscosity
are calculable analytically.  We emphasize that this model is not
particularly well suited for QGP phenomenology; it has no conserved
charge, and also has the peculiar feature of both being nonconformal
\emph{and} having a speed of sound which is independent of temperature.
However, recently it was found that the dynamics of a more complicated
string theory setup were captured by an effective single scalar
Chamblin-Reall background \cite{Bigazzi:2009tc}.  This may be an
indication that there is a nontrivial connection between the
Chamblin-Reall background and nonconformal deformations of more
sophisticated string theory backgrounds.

While the thermodynamics and transport coefficients of this model have
already been studied in the literature \cite{Springer:2008js,
Springer:2009wj, Gubser:2008ny, Gubser:2008yx, Kanitscheider:2009as,
Romatschke:2009kr, Bigazzi:2010ku}, the full bulk spectral density (at
nonzero frequency) has not been previously presented, and we will use the
methods developed in the first part of this paper to do so here.
(This background does not utilize the full power of our methods as it
only contains one scalar field; however, by working in this model we
will be able to perform consistency checks with previous results.) 

Once we have determined the bulk spectral density, we take the
opportunity to investigate the bulk sum rule of
\cite{Romatschke:2009ng} in the context of this model.  We notice an
intriguing fact that for the Chamblin-Reall background, the shear and
bulk spectral densities are simply related at all frequencies.  We
also demonstrate that the bulk sum rule in this model does not agree
with the bulk sum rule written down in \cite{Romatschke:2009ng} and we
explain the reason for the discrepancy.

Recently, a thorough study of the shear correlation function and sum
rule in the Chamblin-Reall model was completed in
\cite{Springer:2010mf}.  The latter half of the present work extends
this analysis to the bulk sector.

The paper is organized as follows.  In Sec. \ref{Background_section}
we detail the multi-scalar gravity dual in which we are working, and
set up the relevant perturbations which need to be analyzed in order
to calculate the two point functions.  In Secs. \ref{action_section}
and \ref{linearized_section}, we generalize the work of
\cite{Gubser:2008sz} by detailing our prescription for calculating the
bulk spectral density, $\pbulk$ in a multi-scalar gravity dual.  This
involves the on-shell gravitational action which is given in
Sec. \ref{action_section} and the solution of relevant equations for
the perturbations given in Sec. \ref{linearized_section}.  Once we
have developed our method for calculating the bulk spectral density,
we apply it to the case of the Chamblin-Reall background in
Sec. \ref{CR_section}.  We provide explicit results for the spectral
density, Euclidean correlation functions, and we also derive a bulk
sum rule in this model.  We summarize our main conclusions and discuss
some possible directions for future work in
Sec. \ref{Conclusion_section}.  Finally, we provide some of the
technical details of the calculations as well as some useful reference
formulae in the appendices.

\section{Background and perturbations}
\label{Background_section}
In this work, we will be interested in the retarded correlation functions
of a strongly coupled, nonconformal field theory,
\begin{equation}
  G_R^{\mu \nu \alpha \beta}(w) \equiv -i \int d^4x\, e^{i w t} \left< \left[T^{\mu \nu} (x),T^{\alpha \beta} (0) \right] \right> \theta(t), \label{GRdef}
\end{equation}
with associated spectral densities
\begin{equation}
  \rho^{\mu \nu \alpha \beta}(w) \equiv -\mbox{Im }G_R^{\mu \nu \alpha \beta}(w) \label{pdef}.
\end{equation}
We will use the labels
\begin{eqnarray}
  G_R^{\rm shear} &\equiv& G_R^{xyxy} \label{GRshear},\\
  G_R^{\rm bulk} &\equiv& \eta_{\mu \nu}\eta_{\alpha \beta} G_R^{\mu \nu \alpha \beta} \label{GRbulk},
\end{eqnarray}
and similarly for the spectral functions.  Here $\eta_{\mu \nu}$
denotes the \emph{four} dimensional Minkowski metric.  Kubo formulas relate
the low energy limit of the spectral functions to the shear and bulk
viscosities: $\eta$ and $\zeta$ respectively
\begin{eqnarray}
  \eta &=& \lim_{w \to 0} \frac{\pshear(w)}{w},\label{etakubo}\\
  \zeta &=& \frac{1}{9}\lim_{w \to 0} \frac{\pbulk(w)}{w}. \label{zetakubo}
\end{eqnarray}
Often in this work, we will be interested in quantities which have their
vacuum (zero-temperature) part subtracted.  We use the symbol $\Delta$ to 
denote this subtraction.  For example, 
\begin{equation}
    \Delta G_R(w) \equiv G_R(w) - G_R(w)_{T=0}.
    \label{Deltadef}
\end{equation}

We will compute the correlation functions of interest using tools provided by
AdS/CFT by working in a dual gravitational theory.  The dual
gravitational theory under consideration is a five-dimensional
multi-scalar theory,
\begin{eqnarray}
	\mathcal{S} &=& \frac{1}{2 \kappa} \int \, d^5 x \sqrt{-g} 
	\left[
	  R - \frac{1}{2} \partial_\mu \phi_a \partial^\mu \phi_a - V(\phi_1,...\phi_n) 
	\right] \nonumber \\ 
	&+& \frac{1}{\kappa} \int d^4 x \sqrt{-\gamma}\, \theta.
\label{action}
\end{eqnarray}
Here, $\kappa$ is related to the five-dimensional Newton's constant,
$\kappa = 8 \pi G_5$.  Summation is implied over the index ``$a$''
which labels each scalar field.  This index runs from $1$ to $n$ with
$n$ denoting the total number of scalar fields.  These scalar fields
in the bulk correspond to operators in the dual field theory.

The second term in (\ref{action}) is a boundary term (the Gibbons-Hawking term).  In
this term, $\gamma_{\mu \nu}$ is the induced metric at the boundary,
$\theta_{\mu \nu}$ is the second fundamental form
\begin{equation}
	\theta_{\mu \nu} = \nabla_\mu \hat{N}_{\nu},
\end{equation}
the unit vector normal to the boundary is denoted by $\hat{N}^{\mu}$,
and $\nabla_\mu$ is the covariant derivative with respect to the background
metric.  

In order to access two point correlation functions in the dual field
theory, one must add perturbations to the background.  
\begin{eqnarray}
  g_{\mu \nu} &\longrightarrow& g_{\mu \nu} + h_{\mu \nu} \\
  \phi_i &\longrightarrow& \Phi_i + \varphi_i.
\end{eqnarray}
The quantities $g_{\mu \nu}$ and $\Phi_i$ are the background fields,
and $h_{\mu \nu}$ and $\varphi_i$ are the perturbations.  The metric
perturbations are generally classified under $SO(2)$ rotations in the
$x_1,x_2$ plane, with the momentum $\vec{q}$ pointing in the $x_3$
direction \cite{Policastro:2002tn,Kovtun:2005ev,Gubser:2008sz}.  The
bulk mode (which is also sometimes called the ``sound mode'' in the
literature) is the scalar mode containing perturbations which are
invariant under such rotations.

In this work, we will be interested in correlation functions at zero
spatial momentum $\vec{q}$, and thus instead of $SO(2)$ rotations, we
have the full $SO(3)$ symmetry.  There are four metric perturbations
which fulfill the requirement of rotational invariance: $h_{00},
h^i_i, h_{05},$ and $ h_{55}$.  Latin indices on the metric components
are assumed to run over the spatial dimensions $i = 1,2,3$.  There are
also $n$ perturbations, one for each scalar field $\varphi_a$.

Not all of these perturbations are physical due to freedom under
diffeomorphisms:
\begin{eqnarray}
  h_{\mu \nu}  &\to& h_{\mu \nu} 
  - \nabla_{\mu} \psi_{\nu} - \nabla_{\nu} \psi_{\mu} \label{hgauge}\\
  \varphi_i &\to& \varphi_i - \psi^{\mu} \partial_{\mu} \Phi_i \label{phigauge}
\end{eqnarray}
for any vector $\psi^{\mu}$.  (Unfortunately, this is often called
``gauge freedom'', though it is not the same as gauge freedom in field
theory).  In \cite{Gubser:2008sz}, only one scalar field was
considered, and the gauge was chosen so that perturbations $\varphi$
and $h_{05}$ vanished.  In this work, we will employ the radial gauge
($h_{05} = h_{55} = 0$) with perturbations $h^0_0$, $h^i_i$, and
$\varphi_a$ nonvanishing.
There are two reasons for this choice of gauge over that of
\cite{Gubser:2008sz}.  First, doing the computation in another gauge
allows for consistency checks between the two calculations.  Secondly,
and more importantly, the choice of the radial gauge allows one to
rather easily generalize the computation to multiple scalar fields.
Such a generalization has not been previously presented, and it is
one of the central results of the present work.

We define the background and perturbations as
\begin{eqnarray}
	ds^2 &=& g_{tt}(z) \left[1 + A(t,z) \right] dt^2 \nonumber \\
	&+& g_{xx}(z)\left[1 + B(t,z)\right]d\mathbf{x}^2 
	+ g_{zz}(z)dz^2, \label{Adef} \\
	\phi_i(t,z) &=& \Phi_i(z) + \varphi_i(t,z).
	\label{varphidef}
\end{eqnarray}
The vector $\vec{x}$ spans the usual 3-spatial coordinates.  The
coordinate $z$ labels the ``extra'' radial dimension; the dual 
four dimensional field theory exists on the boundary at $z = 0$.  The background is
assumed to be static, and all time dependence is contained in the
perturbations $A$,$B$,$\varphi_i$.  

The radial gauge is initially chosen for the purposes of writing down
the relevant equations and on-shell action.  However, it is desirable
to ultimately work in terms of gauge invariant perturbations, which
are invariant under infinitesimal diffeomorphisms
\cite{Kovtun:2005ev,Mas:2007ng,Springer:2008js}.  In
\cite{Springer:2008js}, the more general case of perturbations which
depend on both space and time was examined.  In this more general
case, it was shown that there are $n+1$ gauge invariant variables
where $n$ is the number of scalar fields in the theory.  These
variables were denoted as $\Z_0$ and $\Z_{\Phi i}$.  Here, we are interested
in the special case where the perturbations do not depend on the three
spatial coordinates; in this case, the variable $\Z_0$ vanishes.  We
are left with $n$ gauge invariant variables, one for each scalar
degree of freedom.  These variables have the form:
\begin{equation}
  \mathcal{Z}_{\Phi i}(z) = \varphi_i(z) - \frac{\Phi_i'(z) g_{xx}(z)}{g_{xx}'(z)} B(z).
  \label{Zdef}
\end{equation}
Primes denote derivatives with respect to the radial coordinate $z$.
It is straightforward to check that these variables do not transform
under the diffeomorphisms (\ref{hgauge}) and (\ref{phigauge}).

At first, it is perhaps surprising that the perturbation $A$ (which is
dual to the operator $T^{00}$ in the gauge theory) cannot be put into
a gauge invariant variable.  In fact, this behavior is consistent with
the expected structure of the correlators at zero spatial momentum, as
we will now show.  Because $A$ cannot be put into a gauge invariant
variable, all correlation functions which involve $T^{00}$ vanish (up
to contact terms) in the limit of zero spatial momentum.  The
vanishing of spectral functions involving $T^{00}$ is expected at
$\vec{q} = 0$ on general grounds.  A general correlation function of
components of the energy momentum tensor can be defined to satisfy the
Ward identity \cite{Policastro:2002tn,Kovtun:2005ev}
\begin{equation}
  q_{\mu } G^{\mu \nu \alpha \beta} = \rm{contact\,\,terms}.
  \label{WardIdentity}
\end{equation}
The contact terms which appear on the right hand side of this equation
depend on how the correlation function $G$ is defined.  (Note that $G$
here does not necessarily coincide with the retarded correlation
function $G_R$ defined in (\ref{GRdef})).  Here, the only necessary
detail concerning the contact terms is that they are all \emph{real},
so that regardless of the definition of the correlation function, the
spectral density is the same as that given in (\ref{pdef}).  The Ward
identity implies that the spectral density is transverse:
\begin{equation}
    q_{\mu } \left[\mbox{ Im } G^{\mu \nu \alpha \beta}\right] = -q_{\mu} \rho^{\mu \nu \alpha \beta} = 0 .
\end{equation}
Writing this in components and using the definition of the four-momentum $q^{\mu} = (w,\vec{q})$,
\begin{equation}
  w \rho^{0 \nu \alpha \beta} = q_i \rho^{i \nu \alpha \beta}.  
\end{equation}
Taking the limit where $\vec{q} \to 0$, the right side vanishes
because the correlation functions are nonsingular functions 
at vanishing spatial momentum.\footnote{This is true 
except in some very special cases where excitations at 
$w \neq 0$ do not decay as a function of time \cite{Forster:1975,Yaffe:1992}}  
Then, we see that 
\begin{equation}
  \rho^{0 \nu \alpha \beta}(w, \vec{q} \to  0) = 0.  
\end{equation}
Our calculations in the dual gravity theory will be consistent with
this result.

\section{On-shell action}
\label{action_section}
The prescription for calculating Minkowski space two point functions of a strongly
coupled field theory from gauge/gravity duality was first given in
\cite{Son:2002sd}.  One needs to solve the linearized Einstein
equations for the perturbations, and plug the results back into the
action which has been expended to quadratic order in the
perturbations.  In this section we will expand the action and write it
in terms of the gauge invariant variables.  The main result of this 
section is an expression for the spectral densities in terms of 
the gauge invariant variables.  

Let us assume that we can write our background metric in the coordinate system
\begin{equation}
  ds^2 = g_{xx}(z) \left[ -f(z) dt^2 + d \vec{x}^2 + \frac{dz^2}{f(z)} \right].
  \label{zcoords}
\end{equation}
This coordinate system is only chosen as a calculational aid for the moment.
We shall see that our final results will be valid in any coordinate system.
With the use of the background equations of motion, and the linearized
equations for the perturbations one can write the on-shell action, expanded
to quadratic order in the perturbations, in the form
\begin{eqnarray}
 \mathcal{S}_2 &=& \frac{-V_3}{4\kappa} \int \frac{dw}{2\pi} g_{xx}^{3/2}f \Bigl[ \Z_{\Phi a}(-w,z)\Z_{\Phi a}'(w,z) \Bigr. \nonumber \\
   &+& \Bigl.Q(w,z)^T \xi(z) Q(-w,z)
  \Bigr].
\label{Onshellgaugeinv_text}
\end{eqnarray}
A derivation of this equation is presented in Appendix
\ref{OnShell_app}.  The perturbations depend on $w$ rather than $t$
now because we are working with the Fourier transform.  A summation
over the repeated index $a$ (which takes values from 1 to $n$) is
implied.  Here, $V_3$ denotes the 3-volume (integration over the three
spatial coordinates), $Q$ is a vector made up of perturbations, 
\be
Q(w,z) = \left(
	\begin{array}{c}
	  A(w,z)\\
	  B(w,z)\\
	  \varphi_1(w,z) \\
	  \vdots\\
	  \varphi_n(w,z)
	\end{array}
	\right), 
\ee 
and $\xi$ is a symmetric (2+n) $\times$ (2+n) matrix, the components
of which are unimportant for our present purposes (but are given in
Appendix \ref{OnShell_app}).  This expression for the on-shell action
does \emph{not} contain possible counter terms that are necessary to
regulate UV divergences.  We do not need to consider such counter
terms, because the UV divergences generally only affect the real part
of the on-shell action
\cite{Son:2002sd,Gubser:2008sz,Kaminski:2009dh}.  In this work, we
will only be interested in the spectral functions, which are computed
from the imaginary part of the on-shell action.  In other words,
regularizing counter terms will change the components of the matrix
$\xi$, but this matrix does not contribute to the spectral functions.

The imaginary part of the on-shell
action can be written
\begin{eqnarray}
  \mbox{ Im } \mathcal{S}_{2} 
  &=& \frac{T s V_3}{4 i} \int \frac{dw}{2\pi} \left(\frac{g_{xx}(z)}{g_{xx}(z_h)}\right)^{3/2} \frac{f(z)}{f'(z_h)} \nonumber \\
  &\times& \Bigl[\mathcal{Z}_{\Phi a}'(w,z) \mathcal{Z}_{\Phi a}(-w,z) 
  -(w \to -w)
  \Bigr].
\label{ImOnShellGaugeInv}
\end{eqnarray}
Here the temperature $T$ and entropy density $s$ have been used (details
can be found in Appendix \ref{OnShell_app}).  It
is well known that the imaginary part of the on-shell action is
proportional to a conserved quantity
(see \cite{Son:2002sd,Gubser:2008sz,Kaminski:2009dh}), and thus we can
evaluate the above expression at any value of $z$.  We choose
to evaluate it at the horizon, and make use of the incoming wave
condition
\begin{equation}
  \Z_{\Phi i}'(z_h) (z-z_h) = \frac{-i w}{4 \pi T} \Z_{\Phi i}(z_h)\left[1 + \mathcal{O}(z-z_h)\right],
\end{equation}
and the near-horizon behavior of the metric
\begin{equation}
  \left. \frac{f'(z)}{f(z)}\right|_{z \to z_h} = \frac{1}{z-z_h} + \mathcal{O}(1).
\end{equation}
The result is
\begin{equation}
  -\mbox{ Im } \mathcal{S}_{2} 
  = \frac{s V_3}{8 \pi} \int \frac{w\, dw}{2\pi} \left[\mathcal{Z}_{\Phi a}(w,z_h) \mathcal{Z}_{\Phi a}(-w,z_h)\right], 
  \end{equation}
which can be written
  \begin{eqnarray}
  &&-\mbox{Im } \mathcal{S}_2 = \label{ImSfinal} \\
   && \sum_{a = 1}^n \frac{s V_3}{8 \pi} \int \frac{w\, dw}{2\pi} \Z^0_{\Phi a}(-w)\left[\frac{\mathcal{Z}^h_{\Phi a}(w) \mathcal{Z}^h_{\Phi a}(-w)}{\Z^0_{\Phi a}(-w)\Z^0_{\Phi a}(w)}\right]\Z^0_{\Phi a}(w). \nonumber 
\end{eqnarray}
We have chosen to write it in this form because the quantity in square
brackets is most directly related to the spectral density.  The
superscripts ``0'' and ``h'' denote the boundary and horizon value of
the perturbations at $z \to 0$ and $z \to z_h$.  
Often the boundary values are divergent (though the final answer will
not be).  To be precise, one should introduce a UV regulator
$\epsilon$
\begin{eqnarray}
  \Z^0_{\Phi a} &\equiv& \Z_{\Phi a}(w,\epsilon),\\
  \Z^h_{\Phi a} &\equiv& \Z_{\Phi a}(w,z_h),
\end{eqnarray}
which is assumed to be small.  At the end of the calculation one should take $\epsilon \to 0$. 

We will need to use the fact that due to the definition of
$\Z_{\Phi i}$ (\ref{Zdef}),
\begin{widetext}
\begin{equation}
  \Z^0_{\Phi i}(-w)\Z^0_{\Phi i}(w) = [Q^0(-w)]^T 
   \bordermatrix{
    &&1 & 2 &  ... & 2+i & ... & 2+n \cr \hline \cr
    &&0 & 0 &  ... & 0 & ...& 0 \cr
    &&0 & k_i(\epsilon)^2 & ...& -k_i(\epsilon)& ...& 0 \cr
    &&\vdots & \vdots & \ddots &\vdots&\vdots&\vdots \cr
    &&0 & -k_i(\epsilon) & ... & 1 & ...& 0 \cr
    &&\vdots &\vdots&...&\vdots&\ddots&\vdots \cr
    &&0 & 0 & ... &  0 & ... &0  
}
  Q^0(w)
\,\,\,\,\,(\rm No\,\, sum\,\, over\,\, i),
\label{ZMatrix}
\end{equation}
\end{widetext}
where 
\begin{equation}
  k_i(z) \equiv \frac{\Phi_i'(z) g_{xx}(z)}{g_{xx}'(z)}.
\end{equation}
We have added column labels above the matrix to aid the reader.
The matrix is square, so the same labels apply to the rows.
To be clear, all ellipses ($\dots$) in the above matrix denote
zeros; the nonvanishing elements of this matrix are written
explicitly here.  

The Son and Starinets prescription for the correlation
functions \cite{Son:2002sd}, (see also \cite{Kaminski:2009dh} for a nice treatment of
mixed operators) can be stated as follows.  After writing the on-shell action in
the form
\begin{equation}
  \mbox{Im } \mathcal{S}_2 = \int \frac{dw}{2\pi} [Q^0(-w)]^T M Q^0(w),
\end{equation}
the spectral density matrix is
\begin{equation}
  \rho(w) = -2M.
  \label{SSPrescription}
\end{equation}

For simplicity, we define
\begin{equation}
  R_i(w,\epsilon) \equiv \left[\frac{\mathcal{Z}^h_{\Phi i}(w) \mathcal{Z}^h_{\Phi i}(-w)}{\Z^0_{\Phi i}(-w)\Z^0_{\Phi i}(w)}\right]
  = \frac{ \left|\Z^h_{\Phi i}(w) \right|^2}{\left|\Z^0_{\Phi i}(w) \right|^2}.
\label{Rdef}
\end{equation}
The repeated index $i$ is \emph{not} summed, and the second equality
is due to the fact that $\mathcal{Z}(w)^* = \mathcal{Z}(-w)$.
Combining (\ref{ImSfinal}), (\ref{ZMatrix}),
(\ref{SSPrescription}),and (\ref{Rdef}) we have the main result of
this section:
\begin{widetext}
\begin{equation}
  \rho(w) = \frac{s w}{4 \pi} \lim_{\epsilon \to 0} 
  \bordermatrix{&&\mathcal{O}^A & \mathcal{O}^B & \mathcal{O}^{\Phi_1}& ... & \mathcal{O}^{\Phi_n} \cr \hline \cr
    &&0 & 0 & 0 &...& 0 \cr 
    &&0 & \sum_{a=1}^n k_a(\epsilon)^2 R_a & -k_1(\epsilon)R_1& ...& -k_n(\epsilon)R_n \cr 
    &&0 & -k_1(\epsilon)R_1 & R_1 & ... & 0 \cr
    &&\vdots & \vdots & \vdots & \ddots &\vdots  \cr
    &&0 & -k_n(\epsilon)R_n & 0 & ... & R_n
}.
\end{equation}
\end{widetext}
The ellipses in this matrix do \emph{not} necessarily denote zeros.
To clarify, the entries in the second row and the second column of
this matrix are all nonvanishing except for the zero which is written
explicitly above.  Similarly, the entries on the diagonal are all
nonvanishing, except the upper left corner.  All other entries are
zero.  The column labels are to aid the reader in distinguishing 
the significance of each entry; the matrix is square so the same
labels apply to the rows.  

This matrix contains all information about spectral densities of 
the relevant operators in the dual field theory.  For example, suppose
one is interested in the spectral density of the two point function 
$\left<\mathcal{O}^B \mathcal{O}^{\Phi_1} \right>$ for two operators
$\mathcal{O}^B$ and $\mathcal{O}^{\Phi_1}$ which are dual to the fields
$B$ and $\Phi_1$.  The examination of the above matrix tells us the answer is
\begin{equation}
  \rho^{B \phi_1}(w) = - \lim_{\epsilon \to 0} \frac{s w}{4\pi} k_1(\epsilon) R_1(w,\epsilon).
\end{equation}
Note that all spectral functions involving the operator $\mathcal{O}_A
\sim T^{00}$ vanish.  This is the expected behavior at zero spatial
momentum as explained in Sec. \ref{Background_section}.

In this work, our primary interest is $\pbulk$.  The operator dual to
the perturbation $B$ is $\thalf T^i_i$ (the factor of a half
introduces an extra factor of 4 in the correlation function)
\cite{Policastro:2002tn}.  Thus, we need the matrix element in the
second row, second column:
\begin{equation}
  \pbulk(w) = \frac{s w}{\pi}  \lim_{\epsilon \to 0} \sum_{a=1}^n k_a(\epsilon)^2 R_a(w,\epsilon). 
  \label{rhobulkmulti}
\end{equation}
If one is only interested in the bulk viscosity, one can apply the
Kubo formula (\ref{zetakubo}),
\begin{equation}
  \zeta = \frac{s}{9\pi} \lim_{\epsilon \to 0} \sum_{a=1}^n k_a(\epsilon)^2 R_a(0,\epsilon). 
\end{equation}

It is worth emphasizing that the $k_i$
functions do not change under redefinition of the radial coordinate.
Certainly the bulk spectral density and bulk viscosity which are
physical quantities must not depend on the choice of coordinates in
the gravity dual.  Thus, the $R_a$ functions must also be coordinate
independent (we will be able to see this explicitly in the next
section).  Therefore, this prescription for calculating the spectral
density and bulk viscosity is valid in any background coordinate
system.

In order to evaluate these quantities one must compute the $R_i$
functions by solving the linearized equations for the gauge invariant
fluctuations $\Z_{\Phi i}$.  In the next section we will write down
the relevant equations.

\section{Linearized equations} 
\label{linearized_section} 
In the previous section, we gave results for the bulk spectral function (and
bulk viscosity) in terms of gauge invariant perturbations.  In order to get
an explicit result, one must solve the linearized Einstein equations for the
perturbations.  We will now detail the equations which need to be solved.
\subsection{Theories with multiple scalars}
In \cite{Springer:2008js}, sound mode perturbations were examined for
a general, multi-scalar gravity dual with an arbitrary number of
spatial dimensions denoted by $p$.  The equations for the gauge
invariant variables were given in full generality, assuming both space
and time dependence.  In the general case there is another gauge
invariant variable $\Z_0$ which appears in addition to the $\Z_{\Phi
  i}$ which have already been introduced in the previous section.  The
equation that is relevant for our purposes is\footnote{This is Eq. (66) in
  \cite{Springer:2008js}.  Readers should note a different definition
  of the function $f$ in this paper.  We have re-written the equation
  to correspond with the definitions given in the present work.
}
\begin{eqnarray}
	&\,& \frac{g_{zz}}{\rootg}  \partial_z \left[\rootg g^{zz} \Z_{\Phi i}'\right] 
	-\Z_{\Phi i} g_{zz}\left(w^2 g^{tt} + q^2 g^{xx}\right) \nonumber \\
	&-& \frac{2 k_i}{p} \sum_{a=1}^n  \Z_{\Phi a} \Phi_a' \biggl[\DL \left[\rootg g^{zz} \Phi_a' \right]
	  + \frac{p k_i'}{\alpha k_i}\left(q^2 - \frac{w^2}{f} \right)  \biggr] \nonumber \\ 
	&-&g_{zz}\sum_{a=1}^n \Z_{\Phi a} \frac{\partial^2 V}{\partial \Phi_i \partial \Phi_a}
	+ \frac{2 k_i'}{\alpha \sqrt{f}} \partial_z \left[ \frac{\Z_0}{\sqrt{f}} \right] =0 
\end{eqnarray}
Here, the notation $\DL$ denotes the logarithmic derivative
\begin{equation}
  \DL[Y(z)] = Y'(z)/Y(z),
  \label{DLdef}
\end{equation}
and the quantity $\alpha$ is defined as
\begin{equation}
  	\alpha(z) \equiv q^2 \left((p-1)+ \frac{\DL [g_{tt}]}{\DL[g_{xx}]}\right) - \frac{p w^2}{f}.
\end{equation}
In the current work, we will limit ourselves to the case of zero
spatial momentum $q = 0$, and three spatial dimensions $p = 3$.  When
the spatial momentum of the perturbations is set to zero, the 
gauge invariant variable $\Z_0$ vanishes identically.  Hence, this
equation simplifies to
\begin{eqnarray}
  	&\,& \frac{1}{\rootg}  \partial_z \left[\rootg g^{zz}\Z_{\Phi i}'\right] 
	-w^2 g^{tt} \Z_{\Phi i}  \label{multiZphi} \\
	&-&\sum_{a=1}^n \Z_{\Phi a}\left\{ \frac{\partial^2 V}{\partial \Phi_i \partial \Phi_a}
	+ \frac{2}{3 \rootg }  \partial_z  \left[\rootg g^{zz} \Phi_a' k_i \right] \right\}
	= 0.\nonumber 
\end{eqnarray}
There are $n$ such equations, one for each of the $\Z_{\Phi i}$
variables.  This system of $n$ coupled equations must be solved in
order to determine the bulk spectral density in a generic multi-scalar
gravity dual.  (Usually, the complexity of the equations requires a
numerical solution).  If one is only interested in the bulk viscosity,
one may set $w = 0$ in the above equation.

The equation (\ref{multiZphi}) comes from \cite{Springer:2008js}.  
It is clear from the derivation in that paper, that no special coordinate
system is assumed other than the usual black brane ansatz:
\begin{equation}
  ds^2 = g_{tt}(z)dt^2 + g_{xx}(z)d\vec{x}^2 + g_{zz}(z)dz^2.
\end{equation}
This justifies our discussion in the previous section; equation
(\ref{multiZphi}) is valid in any coordinate system with these
symmetries, and so of course the set of solutions ($R_i$ functions)
is also independent of the coordinate system.  

Let us now summarize the procedure to determine the bulk spectral density
in this general multi-scalar model.  
\begin{enumerate}
  \item
    For the potential $V(\Phi_1...\Phi_n)$ under consideration, one must solve the
    background equations of motion to determine $g_{xx},
    g_{tt}, g_{zz}, \Phi_i ... \Phi_n$.  Once these are known the
    $k_i$ functions are known.  The relevant background equations
    are written in Appendix \ref{equations_app}.

  \item
    Once the background is determined, one must solve the set of
    equations (\ref{multiZphi}) subject to the standard incoming wave
    condition.  This is usually done by inserting the incoming wave
    ansatz
    \begin{equation}
      \Z_{\Phi i}(z) = f(z)^{-i w/ 4 \pi T} Y_i(z)
    \end{equation}
    with the assumption that $Y_i$ is regular at the horizon.  A
    convenient way to do this numerically is to begin numerical
    integration at the horizon and integrate toward the boundary to
    determine the values of $\Z^0_{\Phi i}$.  We refer the reader to
    \cite{Springer:2010mf} as the numerical procedure outlined
    there could be implemented here as well.
  \item
    Once the boundary values $\Z^0_{\Phi i}$ are determined
    numerically, the $R_i$ functions are known, and hence one can use
    the formula (\ref{rhobulkmulti}) to determine the spectral
    density.
\end{enumerate}
This completes our generalization of the work of \cite{Gubser:2008sz} to multiple scalar fields. 

\subsection{Single scalar theories}
If only one scalar field is present, it is possible to completely
remove the potential from the gauge invariant equations, since we can
trade derivatives with respect to $\Phi$ for derivatives with respect
to $z$ using the chain rule.  After doing so and making judicious use
of the background equations of motion (for details, see
Appendix \ref{SingleScalar_app}),  we find the single gauge invariant equation
reduces to the following form.  In the case of one scalar field,
there is only one gauge invariant variable, and so we have dropped all
the $i$ subscripts for simplicity:
\begin{eqnarray}
  &&\frac{1}{\rootg} \partial_z \left[ \rootg g^{zz} \Z_{\Phi}' \right] \nonumber \\
  &-& \Z_{\Phi} \left\{ \frac{1}{\rootg f k} \partial_z \left[\rootg g^{zz} f k' \right] + g^{tt}w^2 \right\}=0 \,.
  \label{Zeqnsingle}
\end{eqnarray}
The associated spectral density is
\begin{eqnarray}
  \pbulk(w) = \frac{sw}{\pi} \lim_{\epsilon \to 0} k(\epsilon)^2 R(w, \epsilon).
  \label{rhobulksingle}
\end{eqnarray}

\section{Chamblin-Reall background}
\label{CR_section}
Despite the relative simplicity of (\ref{Zeqnsingle}), it does not
seem to be possible to solve the equation analytically in general,
even in the limit of vanishing $w$.  However, we are aware of one special case
where certain analytical results are possible.  This case is referred
to as the Chamblin-Reall background, where the scalar potential is a
pure exponential.  The potential is defined as
\begin{eqnarray}
  V(\Phi) = -\frac{6}{L^2} \frac{(2-\delta)}{(1-2\delta)^2} \exp \left\{{\sqrt{\frac{4 \delta}{3}}} \Phi \right\},
  \label{potential2}
\end{eqnarray}
with the numerical factors chosen for later convenience.  The
parameter $L$ has dimensions of length; in the conformal case of
$\delta = 0$ it is the AdS curvature radius.  The metric and
background field are written as follows.
\begin{eqnarray}
  ds^2 &=& b^2(z) \left[ -f(z) dt^2 + d\mathbf{x}^2 + \frac{dz^2}{f(z)} \right] \label{zmetric},\\
  b(z) &=& \left(\frac{L}{z}\right)^{1/(1-2 \delta)},\\
  f(z) &=& 1-\left(\frac{z}{z_h}\right)^{2(2-\delta)/(1-2\delta)},\\
  \Phi (z) &=& -\sqrt{12 \delta} \,\log [b(z)].
\end{eqnarray}
The conformal symmetry breaking parameter $\delta$ ranges from $0 \leq
\delta \leq 1/2$.  It is related to the trace anomaly and speed of
sound as \cite{Kanitscheider:2009as,Springer:2008js},
\begin{equation}
  \delta = \frac{\varepsilon - 3P}{2 \varepsilon} = \frac{1-3v_s^2}{2}
  \label{deltadef}
\end{equation}
where $\varepsilon$, $P$, and $v_s$ denote the energy density, 
pressure, and speed of sound of the dual fluid.  
Note that the Chamblin-Reall background is special, because the function $k(z)$
is independent of $z$.
\subsection{Bulk spectral density and Euclidean correlators}
\label{spectral_Euclidean_subsec}
In this background, the equation of motion for $\Z_\Phi$ becomes
\begin{equation}
  \frac{1}{\rootg g^{zz}} \partial_z \left[ \rootg g^{zz} \Z_{\Phi}' \right] - \Z_{\Phi} g_{zz}g^{tt}w^2 =0.
  \label{ZeqnsingleCR}
\end{equation}
At this point, we notice an interesting coincidence between this
equation for the bulk perturbation, and the case of the \emph{shear}
perturbation.  In \cite{Springer:2010mf}, the shear spectral density
was examined in this model.  In complete analogy with the expressions
above, it was found that
\begin{equation}
  \pshear(w) =\frac{s w}{4 \pi} \left[\frac{H(w,z_h)H(-w,z_h)}{H(-w, \epsilon)H(w, \epsilon)} \right],
\end{equation}
with $H$ being the solution to the equation
\begin{equation}
  \frac{1}{\rootg g^{zz}} \partial_z \left[ \rootg g^{zz} H' \right] - w^2 g_{zz}g^{tt} H = 0.   
\end{equation}
It is clear that the equations for $H$ and for $\Z_\Phi$ are identical
\emph{in this background only}.  (This fact was also recently pointed out
in \cite{Gursoy:2010jh}).  Using the fact that $H=\Z_{\Phi}$, we find
that the bulk and shear spectral densities are very simply related;
their ratio is a constant:
\begin{equation}
  \frac{\pbulk(w)}{\pshear(w)} = 4 \lim_{z \to 0} k(z)^2 = 12 \delta.  
\label{spectralratio}   
\end{equation}
Applying the Kubo formulas to the case at hand, we see that
\begin{equation}
  \frac{\zeta}{\eta} = \frac{4}{3} \delta = 2 \left( \frac{1}{3} - v_s^2 \right),
\end{equation}
which is a well known result (for example, see \cite{Gubser:2008sz,
Springer:2008js}).  One of the novel observations of this work is that
the above ratio between $\zeta$ and $\eta$ is actually a special case
of the more general fact that the full shear and bulk spectral
densities are simply related at all values of $w$.  In Appendix \ref{Gubser_app}, 
we re-derive this result using the methods of \cite{Gubser:2008sz} and show
that both methods give the same answer. 

\begin{figure}
\includegraphics[trim=13mm 0mm 19mm 0mm, clip,width=\columnwidth]{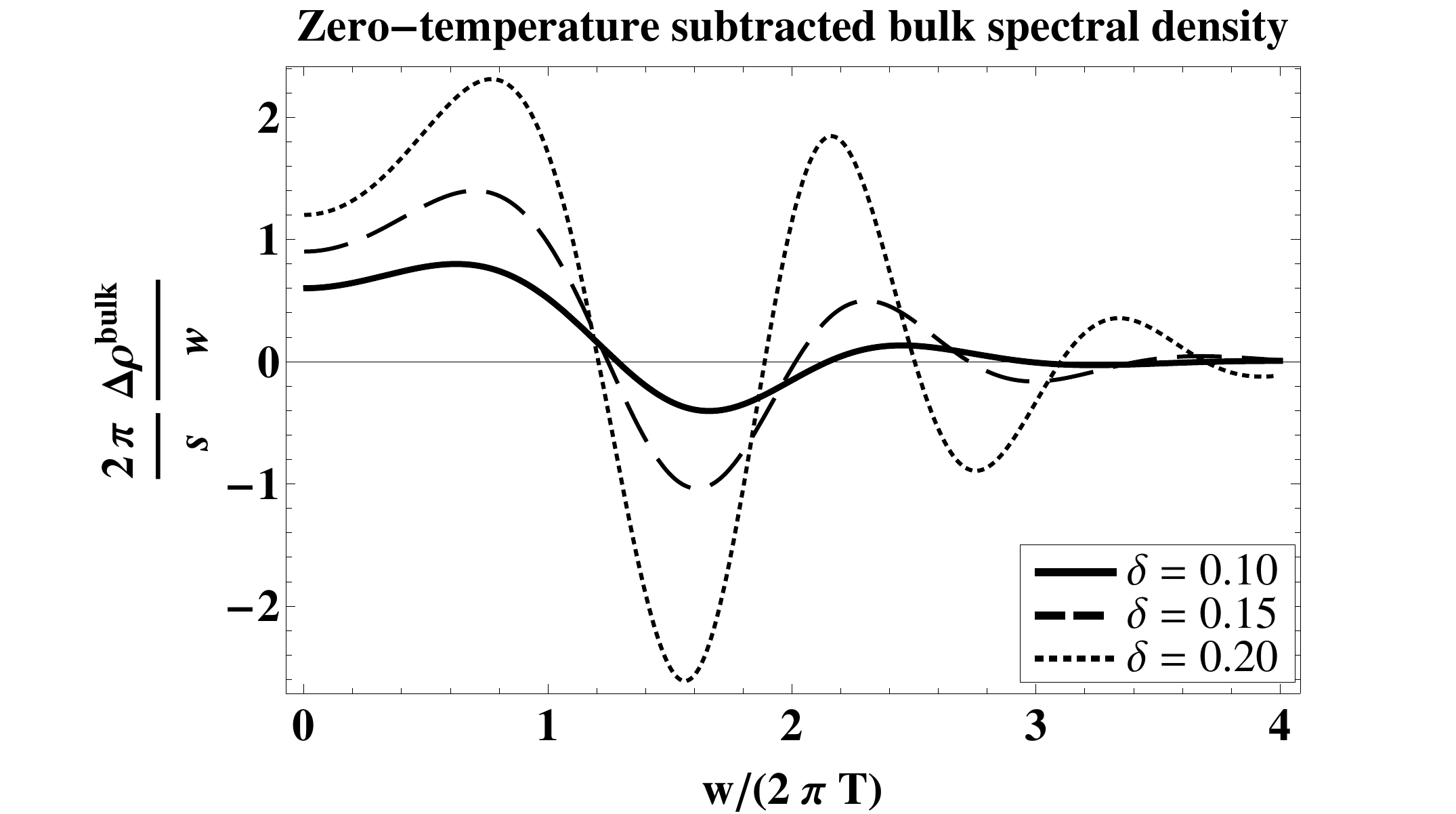}
\caption{ Plots of the zero-temperature subtracted spectral density
versus frequency for several values of $\delta$.  For large $w$, the
spectral density always approaches the zero-temperature result.  Note
that unlike the shear spectral density, the intercept at $w = 0$
increases with $\delta$.  This is a demonstration of the fact that the
bulk viscosity increases with $\delta$.  In a conformal theory,
($\delta = 0$), the spectral density vanishes identically.  }
\label{spectralfigure}
\end{figure}

The shear spectral densities xwere computed in \cite{Springer:2010mf},
and hence we can use the results given there to compute the bulk spectral
densities.  Some results are plotted in Fig. \ref{spectralfigure}.

\begin{figure}
\includegraphics[trim=13mm 0mm 19mm 0mm, clip, width=\columnwidth]{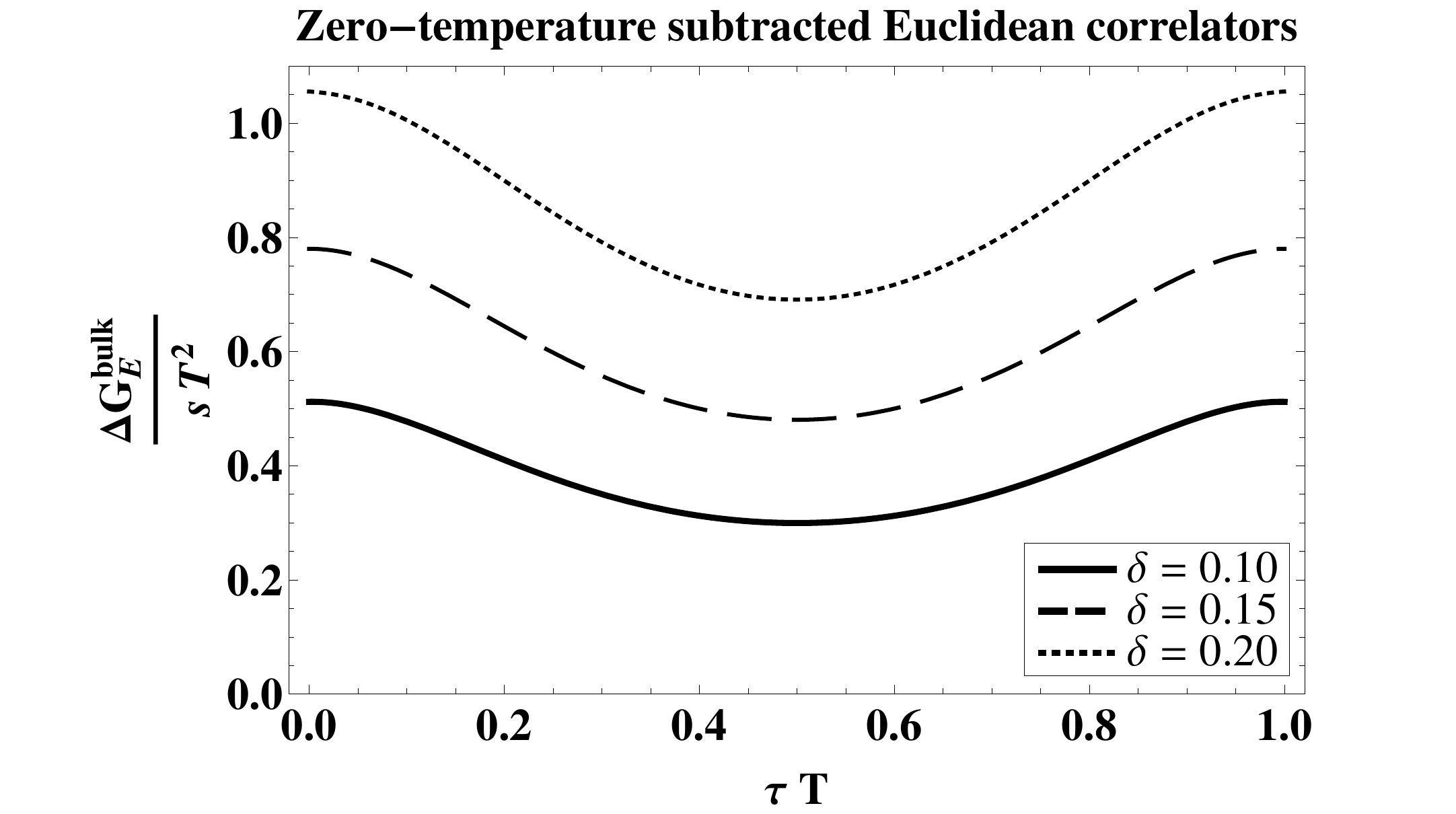}
\caption{ Plot of the zero-temperature subtracted Euclidean
correlation function associated with the bulk spectral density as a
function of the Euclidean time $\tau$ for various values of $\delta$.
In this model the value of $\delta$ (and hence the value of the bulk
viscosity in this model) has a strong effect on the shape and
magnitude of these functions.  In a conformal theory, ($\delta = 0$), 
this function is identically zero.}
\label{GEfig}
\end{figure}

The spectral density is related to the full, Euclidean correlation
function through an integral transform,
\begin{equation}
  G_E(\tau) = \frac{1}{\pi} \int \, dw \rho(w) \frac{\cosh\left[w(\tau - \beta/2) \right]}{\sinh \left[w \beta/2 \right]},
\end{equation}
where $\tau$ is the Euclidean time variable, which has period $\beta
\equiv 1/T$.  Euclidean correlation functions can be computed on the
lattice.  In Fig. \ref{GEfig} we plot this quantity (with the zero
temperature part subtracted) in the bulk channel for the
Chamblin-Reall background.  Despite the
oscillations in the spectral function, the Euclidean correlation
functions turn out to be smooth.

\subsection{Bulk sum rule}
With the spectral density in hand, it is a simple matter to
investigate the bulk sum rule of \cite{Romatschke:2009ng}.  The shear
sum rule in this background was already studied in
\cite{Springer:2010mf}, with the result
\begin{equation}
  \frac{3}{10-8\delta}(\varepsilon + P) = \int_0^{\infty} \frac{dw}{w} \left[\rho^{\rm shear}(w) - \rho^{\rm shear}_{\rm T=0}(w) \right].   
\end{equation}
Multiplying both sides by $12 \delta$, and using
(\ref{spectralratio}), we find the result for the bulk channel
\begin{equation}
  \frac{36 \delta}{10-8\delta}(\varepsilon + P)  = \int_0^{\infty} \frac{dw}{w} \left[\rho^{\rm bulk}(w) - \rho^{\rm bulk}_{\rm T=0}(w) \right].   
\end{equation}
In \cite{Romatschke:2009ng}, the left side of the above sum rule was
derived for Yang-Mills theories, and was found to take on the
value
\begin{equation}
  3(\varepsilon + P)(1-3 v_s^2) - 4 \left (\varepsilon - 3P\right). 
\end{equation}
If we naively evaluate this quantity in the Chamblin-Reall metric using (\ref{deltadef}), we
find
\begin{equation}
  3(\varepsilon + P)(1-3 v_s^2) - 4 \left (\varepsilon - 3P \right) = \frac{-6 \delta^2}{2-\delta}(\varepsilon + P). 
\end{equation}
Clearly, there are differences between the bulk sum rule of
\cite{Romatschke:2009ng} and the bulk sum rule for the Chamblin-Reall
metric (both the sign and the leading power of  $\delta$ are
different).

What is the nature of this difference?  The derivation of the sum rule
\cite{Romatschke:2009ng} employs asymptotic freedom, and so we should 
examine the beta function in the Chamblin-Reall background to see
whether this model meets the necessary requirements.

\subsection{Beta function and trace anomaly}
One can show that in single scalar gravity duals, the trace anomaly is
consistent with that of QCD,
\begin{equation}
  	\left<\mbox{Tr}(F^2) \right> = 4 \left( \varepsilon - 3P \right) \left(\frac{\lambda^2_t}{\beta(\lambda_t)} \right),
\end{equation}
provided that we identify \cite{Gursoy:2008za}
\begin{eqnarray}
  \left<\mbox{Tr}(F^2) \right> &=& \lambda_t \frac{8\sqrt{2}}{\sqrt{3}} \left< \mathcal{O}_{\Phi} \right>, \\
  \frac{\beta(\lambda_t)}{\lambda_t} &=& \sqrt{\frac{3}{8}} b \frac{d \Phi}{d b}.
\end{eqnarray}
Here $\lambda_t = g^2 N_c$ is the t'Hooft coupling, $F^2$ is the
square of the field strength tensor $F_{\mu \nu}^a$, and $b$ is
defined from the metric as written in the coordinate system
(\ref{zmetric}).  The field theory operator dual to the scalar field
$\Phi$ is denoted as $\mathcal{O}_{\Phi}$.  The t'Hooft coupling is
usually identified with a scaled version of the dilaton
\begin{equation}
  e^{\sqrt{\frac{3}{8}}\Phi} = c_0 \lambda_t
\end{equation}
with $c_0$ being an unknown proportionality constant which will not enter our equations.  Note that
\begin{equation}
  \frac{\beta(\lambda_t)}{\lambda_t} = \sqrt{\frac{3}{8}} b(z) \frac{d \Phi}{d z}\frac{dz}{db} 
  = \sqrt{\frac{3}{2}} \frac{\Phi'(z)g_{xx}(z)}{g_{xx}'(z)} = \sqrt{\frac{3}{2}} k(z).  
\end{equation}
In order to get an explicit result for $\beta$ in the general case,
one would have to invert $\lambda_t(z)$ and substitute it above to get
$k(z(\lambda_t))$.  In the case of the Chamblin-Reall metric, this is
unnecessary because $k$ is a pure constant.  In this background, then
\begin{equation}
  \beta(\lambda_t) = - 3\lambda_t \sqrt{\frac{\delta}{2}}. 
\end{equation}
The beta function is negative, and hence this model possesses
asymptotic freedom.   There must be another reason why the sum
rule in this model does not agree with that of
\cite{Romatschke:2009ng}.

The left side of the sum rule is, more generally,
\begin{equation}
  \Delta G_R^{\rm bulk}(w = i \infty) - \Delta G_R^{\rm bulk}(w = 0),
\end{equation}
where we use the symbol $\Delta$ to denote a zero-temperature
subtraction as in (\ref{Deltadef}).
The low energy piece (at $w = 0$) provides a universal contribution
from hydrodynamics, so this is not the source of the difference.  The
crucial difference is the source of conformal symmetry breaking in
each model.  In gauge theories, the conformal symmetry is broken by
the running of the coupling.  If the theory is asymptotically free, at
large energy scales the coupling vanishes due to asymptotic freedom,
and hence the high energy term (at $w = i \infty$) vanishes.  In other
words, conformal symmetry is restored at high energies or temperatures
in such gauge theories.

In the Chamblin-Reall theory, there is a hard breaking of conformal
symmetry due to a parameter in the Lagrangian.  This parameter does
not run with energy scale and hence at large energies or temperatures,
conformal symmetry is \emph{not} restored, and thus $\Delta G_R(w =
i\infty)$ does not vanish in general.  Thus, the different form of the
left side of the sum rule in this theory is a consequence of the fact
that the conformal symmetry breaking is present for all temperatures.

Despite the fact that the left side of the sum rule appears different
in each theory, we can find some similarity by writing the left side
in terms of the beta function.  In what follows, we will assume that
the conformal symmetry breaking is small $(\varepsilon - 3P) \ll
\varepsilon,P$.

\subsubsection{Yang-Mills theory}
For gauge theories, the region of small conformal symmetry breaking
occurs at weak coupling.  The pressure
can be computed as
\begin{equation}
  P(T) = T^4 \left(A + B g^2 + \mathcal{O}(g^3)\right).
\end{equation}
using $s = P'(T)$, $\epsilon + P = T s$ and $ v_s^2 =
P'(T)/\varepsilon'(T)$, one finds that, to lowest order in the
coupling constant $g$, the left hand side of the Romastschke-Son sum rule
is
\begin{equation}
  3(\varepsilon+P)(1-3v_s^2) -4 (\varepsilon - 3P) \approx B T^5 \frac{d \beta(g^2)}{dT},
\end{equation}
Where 
\begin{equation}
  \beta(g^2) = T \frac{d g^2}{dT}.
\end{equation}
The beta function is computed as
\begin{equation}
  \beta(g^2) = -b_0 g^4 + b_1 g^6 + ...
\end{equation}
Again, to leading order, 
\begin{equation}
  T \frac{d \beta(g^2)}{dT} = -2 b_0 g^2 \beta(g^2) = 2 \left(\frac{\beta(g^2)}{g}\right)^2.
\end{equation}
In all, then the left side of the sum rule becomes (to leading order in the conformal symmetry breaking
parameter, $g$), 
\begin{equation}
  3(\varepsilon+P)(1-3v_s^2) -4 (\varepsilon - 3P) \approx \frac{2B}{A} P_0 \left(\frac{\beta(g^2)}{g}\right)^2
\end{equation}
where $P_0$ denotes the pressure in the conformal limit $P_0 = A T^4$.
Inserting factors of $N_c$ and using the fact that the t'Hooft
coupling is $\lambda_t = g^2 N_c$, we find
\begin{equation}
  3(\varepsilon+P)(1-3v_s^2) -4 (\varepsilon - 3P) \approx \frac{2B \lambda_t}{A N_c} P_0 \left(\frac{\beta(\lambda_t)}{\lambda_t}\right)^2. 
  \label{LHSYM}
\end{equation}
 
\subsubsection{Chamblin-Reall background}
For the Chamblin-Reall background, the conformal symmetry breaking parameter is $\delta$.  
The left side of the sum rule is to lowest order:
\begin{equation}
  \frac{18}{5} \delta \times 4P_0 + \mathcal{O}(\delta^2),
\end{equation}
Here, again, $P_0$ denotes the pressure in the conformal limit $P_0 =
P(\delta \to 0)$, and we have used the fact that $\varepsilon = 3P$ in
the conformal limit.  Employing the use of the beta function
\begin{equation}
  \left(\frac{\beta(\lambda_t)}{\lambda_t} \right)^2 = \frac{9}{2} \delta,
\end{equation}
we find the left side of the sum rule can be written 
\begin{equation}
  \frac{16}{5} P_0 \left(\frac{\beta(\lambda_t)}{\lambda_t} \right)^2.
  \label{LHSCR}
\end{equation}
Comparing (\ref{LHSCR}) to (\ref{LHSYM}), we note that when the conformal symmetry breaking
is small, the bulk sum rule is well approximated in both theories as
\begin{equation}
  C P_0 \left(\frac{\beta(\lambda_t)}{\lambda_t} \right)^2 
  \approx \int_0^{\infty} \frac{dw}{w} \left[\rho^{\rm bulk}(w) - \rho^{\rm bulk}_{\rm T=0}(w) \right].
\end{equation}
This equation is approximate because we have only included the leading
order term in the deviation from conformality on the left side.  While
the functional form is similar in both theories, $C$ is theory
dependent.  For the Chamblin-Reall background we have shown $C =
16/5$; for pure Yang-Mills theory \cite{Kapusta2006}, it is $C =
-5\lambda_t/8 \pi^2$.  It would be interesting to see whether this
dependence holds in other theories beyond those considered here.  It
is interesting that $C$ contains a different dependence on the coupling
in each theory; this is a reflection of the fact that to lowest order
the beta function goes as $\lambda_t$ in Chamblin-Reall, but as
$\lambda^2_t$ in Yang-Mills theory.

In the theories we have considered, the left hand side of the bulk sum
rule has the same dependence (on the conformal symmetry breaking
parameter) as the bulk viscosity.  In fact, it has been argued on
general grounds that the bulk viscosity should be proportional to
\emph{two} factors of the beta function in \cite{Arnold:2006fz}.
Despite the fact that the sum rule (and bulk viscosity) both depend on
two powers of the beta function in both theories, the dependence on
$(\varepsilon - 3P)$ is different.  By looking at the Kubo formula for
the bulk viscosity one might expect the dependence $\zeta \sim
(\varepsilon - 3P)^2$. Here, we see this is not generally true.  The quantity 
$\varepsilon - 3P$ is not necessarily linearly proportional to the beta
function.

\section{Conclusion}
\label{Conclusion_section}
The main results of this work were presented in two parts.  First, we
presented a prescription for calculating the bulk spectral density (and
hence the bulk viscosity) for a general, five-dimensional multi-scalar
gravity dual theory.  In this way, we have generalized the results of
\cite{Gubser:2008sz} to multiple scalar fields.  In so doing, we have
calculated the on-shell gravitational action relevant for the
computation of bulk mode two point correlation functions.  We have
also given explicitly the set of gauge invariant equations which need
to be solved for a given background.  As such, our main results for
this part are summarized at the end of Sec. \ref{linearized_section}.

There are many gravitational dual theories on the market today.  For
this reason, it would be useful to have a definite prescription which
allows for the calculation of the bulk spectral function which works
for a large class of gravity duals.  We have taken a step in
this direction by including multiple scalar fields.  Also, as
mentioned in the introduction, more than one scalar field often
appears when deriving a five-dimensional effective action for a string
theory setup.  Hence, the methods we have developed here could aid future computations in 
backgrounds like those of \cite{Benincasa:2005iv, Aharony:2005zr}.  

In the second part of the paper, we used the methods derived in the
first part to compute the bulk spectral function (at vanishing spatial
momentum $\vec{q}$) in the Chamblin-Reall background.  We noticed the
intriguing fact that in this model the ratio of the bulk to shear
spectral densities is a pure constant.  We also derived a sum rule for
the bulk channel and show that it is different from the sum rule of
Romatschke and Son \cite{Romatschke:2009ng}.  The reason for this
difference is not due to asymptotic freedom, but rather due to the
``hard'' nature of the conformal symmetry breaking in the Chamblin-Reall
background.  We have also shown that (to leading order in the
conformal symmetry breaking) the left side of the sum rule in both
Yang-Mills theory and in the Chamblin-Reall background is proportional
to two factors of the beta function.

When one attempts to extract transport coefficients from the lattice,
one typically needs to make an ansatz for the spectral
function.  The results from AdS/CFT computations like the one
presented here could be used in this process (though, as we mentioned
previously, it is necessary to perform the computation in a theory
which shares more phenomenological features with QCD). 

There are many possible directions for future work.  It would be
useful to further generalize our prescription by including other
matter fields beyond the simple scalars included here.  The inclusion
of vector fields allows one to study holographic models at nonzero
chemical potential.  One could also try to generalize our methods to
an arbitrary number of dimensions.  Furthermore, it would be
interesting to compute the correlation functions at nonvanishing
spatial momentum.  In this way, one could have a complete
generalization of the known conformal results of \cite{Kovtun:2006pf}.
We hope to address some of these issues in future publications.

\section*{Acknowledgments}
We thank Aldo Cotrone, Paul Romatschke, Dam Son and Mikhail Stephanov for
helpful discussions.  This work was supported by the Natural Sciences
and Engineering Research Council of Canada.  

\appendix

\section{Equations of motion}
\label{equations_app}
In this section we write down the relevant background Einstein equations
and the linearized Einstein equations for the perturbations.  
\subsection{Background equations}
There are $3+n$ independent background equations, corresponding to
three metric components $g_{tt},g_{xx},g_{zz}$ and $n$ scalar fields
$\Phi_i$.  For later convenience, we write these equations as the
vanishing of the quantities
\begin{equation}
  \E0_1 = \E0_2 = \E0_3 = \E0_{\Phi i} = 0
\end{equation}
with the definitions
\begin{eqnarray}
  \E0_1 &\equiv& V(\Phi_1,...\Phi_n) + \frac{3}{2\rootg} \partial_z \left( \rootg g^{zz} \DL[g_{xx}] \right) \label{E01def}\\
  \E0_2 &\equiv& \Phi_a' \Phi_a' - \frac{3}{2} \DL[g_{xx}] \DL \left[ \frac{g_{zz}f}{\DL[g_{xx}]^2} \right] \\
  \E0_3 &\equiv& \frac{1}{2\rootg} \partial_z \left( \rootg g^{zz} \DL[f] \right)\\
  \E0_{\Phi i} &=& \frac{\partial V}{\partial \Phi_i} - \frac{1}{\rootg} \partial_z \left( \rootg g^{zz} \Phi_i' \right) \label{E04def}.
\end{eqnarray}
The superscript $(0)$ denotes that these are background (zeroth order)
equations.  The logarithmic derivative $\DL$ was introduced in
(\ref{DLdef}).  A summation over the repeated index $a$ is implied.

\subsection{Linearized equations}
There are 2+n independent linearized Einstein equations (first order
in the perturbations) corresponding to the two metric perturbations
$A,B$ and the $n$ scalar perturbations $\varphi_i$.  In
\cite{Springer:2008js}, the linearized perturbations for this system
were given in the more general case of $\vec{q} \neq 0$; in the general case
there are 4 metric perturbations denoted as $A,B,C,D$.  One can get
the equations necessary for the current work by examining the
equations in \cite{Springer:2008js} in the limit of $q = 0, D = 0, B =C$. 
We will write the equations for the Fourier transformed
perturbations, such as
\begin{equation}
  A(t,z) = \int \, \frac{dw}{2\pi} A(w,z) e^{-i w t}, 
\end{equation}
with the shorthand $A_w$, $B_w$, $\varphi_{i,w}$ denoting $A(w,z)$, $B(w,z)$, $\varphi_i(w,z)$.  
As in the previous subsection we write these equations as the vanishing of the quantities
\begin{equation}
  \e1_1 = \e1_2 = \e1_{\Phi i} = 0
\end{equation}
with the definitions
\begin{equation}
  \e1_1 \equiv 3 \sqrt{f} \partial_z \left[ \frac{B_w}{\sqrt{f}} \right] + \Phi_a' \varphi_{a,w} \,, \label{EOM4}
\end{equation}
\begin{eqnarray}
  \e1_2 &\equiv& \DL[g_{xx}^3]A_w' + 3\DL[g_{xx}^2g_{tt}]B_w' - 2 \Phi_a' \varphi_{a,w}' 
  \nonumber \\
  &+& \frac{2 \varphi_{a_,w}}{\rootg g^{zz}} \partial_z \left[ \rootg g^{zz} \Phi_a' \right] - 6w^2 g_{zz}g^{tt}B_w \,,
  \label{EOM5}
\end{eqnarray}
\begin{eqnarray}
  \e1_{\Phi i} &\equiv& \frac{1}{\rootg g^{zz}} \partial_z \left[ \rootg g^{zz} \varphi_{i,w}'\right] 
  + \frac{\Phi_i'}{2} \left(A_w'+3B_w' \right) \nonumber \\
  &-& g_{zz} \frac{\partial^2 V}{\partial \Phi_i \partial \Phi_a} \varphi_{a,w} 
  -w^2 g_{zz}g^{tt} \varphi_{i,w} \label{LinearScalarEOM}\,.
\end{eqnarray}
Summation over the repeated index $a$ is implied, and we have used the background equations
to remove any reference to the potential in (\ref{EOM5})

\section{Derivation of on-shell action}
\label{OnShell_app}
In this section, we provide a calculation of the on-shell action
in terms of gauge invariant variables, leading to the expressions
(\ref{Onshellgaugeinv_text}) and (\ref{ImOnShellGaugeInv}).  Throughout this section we will assume that the
metric is written in the coordinate system (\ref{zcoords}).  Because it is
sometimes cumbersome to write down square matrices of arbitrary rank, we will 
write any necessary matrices for \emph{two} scalar fields.  It will be
immediately obvious how to generalize our final result for more than two 
scalar fields.  

To begin, we start with the action (\ref{action}) and the
perturbations defined in (\ref{Adef}),(\ref{varphidef}).  Next, one
expands the action to second order in the perturbations.
\subsection{Bulk term}
We first focus on the ``bulk term'': the first term in (\ref{action}).  The part
quadratic in the perturbations can be written

\begin{eqnarray}
  \mathcal{S}^{\rm bulk}_2 &=& \frac{1}{2\kappa}\int d^5x \left[ \Q^T M \Q\right. + (\Q')^T M_{zz} \Q' 
    + (\Q')^T M_z \Q 
    \nonumber \\ &+& (\Q'')^T M_{2z} \Q 
    + \left. (\ddot{\Q})^T M_{2t} \Q + (\dot{\Q})^T M_{tt}\dot{\Q}\right].
\end{eqnarray}
Here, $\Q$ is the matrix of perturbations
  \be
	\Q(t,z) = \left(
	\begin{array}{c}
	  A(t,z)\\
	  B(t,z)\\
	  \varphi_1(t,z) \\
	  \vdots\\
	  \varphi_n(t,z)
	\end{array}
	\right),
\end{equation}
the dot denotes derivatives with respect to time, and the prime
denotes derivatives with respect to $z$.  And we have introduced
several $2+n$ by $2+n$ matrices:$M, M_{zz}$, etc.  We will not write
down the components of all these matrices explicitly here, as only a
fraction of these components enter into our final result.

We will work with the Fourier modes, by inserting
\begin{equation}
  \Q(t,z) = \int \frac{dw}{2\pi} Q(w,z) e^{-i w t}.  
\end{equation}
And we will employ the shorthand 
\begin{equation}
  Q_w = Q(w,z).
\end{equation}
The on-shell action becomes
\begin{eqnarray}
  \mathcal{S}_2^{\rm bulk} &=& \frac{1}{4 \kappa} \int d^3 x dz \frac{dw}{2 \pi} \,\biggl\{2 (Q_w')^T M_{zz} Q_{-w}' 
  \nonumber \\ &+& (Q_w')^T M_z Q_{-w} +  Q_{w}^T M^T_z Q_{-w}'
  \nonumber \\
  &+& \biggl. (Q_{w}'')^T M_{2z} Q_{-w} + Q_{w}^T M_{2z}^T Q_{-w}''+Q_{w}^T M_c Q_{-w} \biggr\} \nonumber 
\end{eqnarray}
with $M_c$ defined as
\begin{equation}
  M_c \equiv 2 M + w^2 \left(M_{tt} + M_{tt}^T - M_{2t} - M_{2t}^T \right).
\end{equation}
Note that because of the integration over all $w$, (from $-\infty$ to
$\infty$) we have taken the opportunity to symmetrize the entire
expression in $w \to -w$.  Terms which are odd in $w$
will drop out under the $w$ integration.  
Integrating by parts to get an action which depends only on first derivatives, one has
\begin{eqnarray}
  \mathcal{S}_2^{\rm bulk} &=& \frac{V_3}{4 \kappa} \int dz \frac{dw}{2\pi} \, \left[(Q_w')^T M_a Q_{-w}'\right. 
    + (Q_{w}')^TM_b Q_{-w} \nonumber \\
    &+& Q_{w}^T M_b^T Q_{-w}' 
    + \left.Q_w^T M_c Q_{-w} + \partial_z \Delta(w,z) \right]
\end{eqnarray}
with
\begin{eqnarray}
  M_a(z) &\equiv& 2 M_{zz} - M_{2z} - M_{2z}^T \\
  M_b(z) &\equiv& M_z - M_{2z}'\\
  \Delta(w,z) &\equiv& (Q_w')^T M_{2z} Q_{-w} + Q_{w}^T M_{2z}^T Q_{-w},
\end{eqnarray}
and we have introduced the
result of the integration over $d^3x$ as $V_3$.
One can now employ the Euler-Lagrange equations to get the equations of
motion for the perturbations.  In doing so, one should treat $Q_w$ and
$Q_{-w}$ as separate functions.  (The reason is that one has in mind a
complex perturbation so that $Q_{-w} \sim Q^*_{w}$
\cite{Gubser:2008sz}.)  One can show that upon application of the
equations of motion, this entire term in the on-shell action reduces
to a total derivative term.  This procedure has been explained many
times in the literature (e.g. \cite{Gubser:2008sz, Buchel:2004di,
Kaminski:2009dh}) hence we will not go through all of the details.
The result of this procedure is
\begin{eqnarray}
  \mathcal{S}_2^{\rm bulk} &=& \frac{V_3}{4 \kappa} \int \frac{dw}{2\pi} \Bigl[(Q_{w}')^T(M_a + M_{2z})Q_{-w} \Bigr. \nonumber \\
    &+& \Bigl. Q_w^T M_{2z}^T Q_{-w}' 
    + Q_{w}^T M_b^T Q_{-w} \Bigr]_{z=z_B}.
\end{eqnarray}
Here $z=z_B$ is a symbolic way of noting that the above should be
evaluated at the boundary of the space-time.
\subsection{Gibbons-Hawking term}
We now must add the second term of (\ref{action}), which is itself a pure boundary term.   
After expanding this term to second order in the perturbations, we find
\begin{equation}
  \mathcal{S}_2^{GH} = \frac{1}{2 \kappa} \int (\Q')^T M_{GHa}\Q + \Q^T M_{GHb} \Q.
\end{equation}
with two new matrices $M_{GHa}$, $M_{GHb}$.  Repeating the arguments
of the previous section, by inserting the Fourier modes and
symmetrizing in $w$, we find the Gibbons-Hawking contribution
\begin{eqnarray}
  \mathcal{S}_2^{GH} &=& \frac{V_3}{4 \kappa} \int \frac{dw}{2\pi} \Bigl[(Q_w')^T M_{GHa}Q_{-w} + Q_{w}^T M_{GHa}^T Q_{-w}'
    \Bigr.\nonumber \\
    &+& \Bigl. Q_w^T (M_{GHb}+M_{GHb}^T) Q_{-w} \Bigl]_{z=z_B}.
\end{eqnarray}
By an explicit computation, we find that both $M_{2z}$ and $M_{GHa}$ are symmetric, and that
\begin{equation}
  M_{GHa} = -M_{2z}.  
  \label{MGHprop}
\end{equation}
Indeed, this should come as no surprise, since the purpose of the Gibbons-Hawking term is to 
cancel off the boundary contribution due to the integration by parts. 
\subsection{Total}
 Adding together the bulk and Gibbons-Hawking terms, and using (\ref{MGHprop}), we find
\begin{eqnarray}
  \mathcal{S}_2 &=& \mathcal{S}_2^{\rm bulk} + \mathcal{S}_2^{GH} \\
  &=& \frac{V_3}{4 \kappa} \int \frac{dw}{2\pi} (Q_w')^T M_a Q_{-w} + Q_{w}^T(M_b^T + 2 M_{GHb})Q_{-w} \nonumber.
\end{eqnarray}
We are now ready to write down the forms of the matrices appearing here explicitly.  For the case of
\emph{two} scalar fields, the matrices are 4 $\times$ 4, and are given by
\begin{equation}
  M_a =  \frac{f g_{xx}^{3/2}}{2} 
  \bordermatrix{&& & & &  \cr
    &&0 & 3 & 0 & 0 \cr
    &&3 & 6 & 0 & 0 \cr
    &&0 & 0 & -2 & 0 \cr
    &&0 & 0 & 0 & -2
}
\end{equation}
and
\begin{eqnarray}
  && M_b^T + 2 M_{GHb} = \label{Mbfinal} \\
  &&\frac{f g_{xx}^{3/2}}{4} \left(
  \begin{array}{cccc}
    -3 \DL[g_{xx}] & 3\DL[f g_{xx}^3] & -2\Phi_1' & -2 \Phi_2' \\
    9 \DL[g_{xx}] & 3 \DL[f g_{xx}^3] & -6 \Phi_1' & -6 \Phi_2' \\
    0 & 0 & 0 & 0 \\
    0 & 0 & 0 & 0
  \end{array}
  \right) \nonumber 
\end{eqnarray}
It should be evident how to generalize these matrices for more than
two scalars.  Regardless of the number of scalars, the $2 \times 2$
submatrix in the upper left corner always remains the same, since
these entries correspond to the metric perturbations $A$ and $B$.  For
more scalar fields, $M_a$ simply has more entries of $-2$ along the
diagonal.  Similarly, (\ref{Mbfinal}) will in general only have
non-zero entries in the first two rows; one adds entries
$-2\Phi_3'...-2\Phi_n'$ to the first row, and $-6\Phi_3'...-6 \Phi_n'$ to
the second row.
\\ 
\subsection{Gauge invariant form}
We would like to write the action involving the gauge-invariant variables $\Z_{\Phi i}$.
Our strategy is to employ the equations of motion to remove all derivatives except for
$\Z_{\Phi i}'$.  Specifically, we first use (\ref{EOM5}) to remove all instances of $A'$,
then remove all $\varphi'$ in favor of $\Z_{\Phi i}'$ and $B'$.  Finally, use (\ref{EOM4}) 
to remove all remaining $B'$ terms.  The result is
\begin{eqnarray}
  &&\mathcal{S}_2 - \frac{V_3}{8 \kappa} \int \frac{dw}{2\pi} \frac{g_{xx}^{5/2}f}{g_{xx}'} 
  \Biggl\{
  B_{-w}\e1_2 -  \frac{2\Z_{\Phi a}(-w,z)\Phi_a'}{3} \e1_1 \Biggr. \nonumber \\
  &-& \Biggl. \Bigl[ \DL[g_{xx}f]B_{-w} - \DL[g_{xx}]A_{-w} \Bigr]\e1_1
  \Biggr\} \label{OnShellGaugeinv} \\
  &=& \frac{-V_3}{4\kappa} \int \frac{dw}{2\pi} g_{xx}^{3/2}f \left\{ \Z_{\Phi a}(-w,z)\Z_{\Phi a}'(w,z) + Q_{w}^T \xi Q_{-w}
  \right\}\nonumber
\end{eqnarray}
The terms on the left side vanish on-shell due to the equations of motion $\e1_1 = \e1_2 = 0$.  
The term without derivatives (the contact term) contains the matrix $\xi$, the components of which can be written (in the case
of two scalar fields):
\begin{widetext}
\begin{equation}
  \xi(z) \equiv 
  \left(
  \begin{array}{cccc}
    \frac{3}{4} \DL[g_{xx}] & -\frac{3}{4} \DL[f g_{xx}^3] & \frac{\Phi_1'}{2} & \frac{\Phi_2'}{2} \\
     -\frac{3}{4} \DL[f g_{xx}^3] & \frac{3}{4}\chi(z) & k_1 \DL[k_1 g_{xx}^{3/2} \sqrt{f}] & k_2 \DL[k_2 g_{xx}^{3/2} \sqrt{f}] \\
     \frac{\Phi_1'}{2} & k_1 \DL[k_1 g_{xx}^{3/2} \sqrt{f}] & -\frac{\Phi_1^2}{3 \DL[g_{xx}]} & -\frac{\Phi_1'\Phi_2'}{3 \DL[g_{xx}]} \\
     \frac{\Phi_2'}{2} & k_2 \DL[k_2 g_{xx}^{3/2} \sqrt{f}] & -\frac{\Phi_1'\Phi_2'}{3 \DL[g_{xx}]}& -\frac{\Phi_1^2}{3 \DL[g_{xx}]}
    \end{array}
  \right) 
  - \left(\begin{array}{cccc}
  0&0&0&0 \\
  0 & \frac{1}{2f} \partial_z \left(\frac{f \E0_2}{\DL[g_{xx}]^2}\right) & 0 & 0 \\
  0 & -\frac{k_1}{3\DL[g_{xx}]} \E0_2 & 0 & 0 \\
  0 & -\frac{k_2}{3\DL[g_{xx}]} \E0_2 & 0 & 0
  \end{array}\right)
\end{equation}
\end{widetext}
The second matrix here is proportional to the background equations of motion, and thus vanishes on-shell.  
The function $\chi(z)$ is defined as
\begin{equation}
  \chi(z) \equiv 
    \frac{\DL[f ]^2}{\DL[g_{xx} ]}
  - \DL\left[f g_{xx} ^3\right]
  \\
  - 
    \frac{2}{f }\partial_z\left[f \partial_z\left(\frac{g_{xx} }{g_{xx}' }\right)\right]+\frac{4 w^2g_{xx} }{f ^2 g'_{xx}}.
\end{equation}
It is clear that $\xi$ is a symmetric matrix (up to terms which vanish
upon application of the equations of motion).  For this reason, it
does not enter in the computation of the spectral density which we
consider in this work.  The matrix $\xi$ may be useful in future work,
if one is interested in computing the $\emph{real}$ part of the correlation functions.  
Such a computation would also require extra counter terms to renormalize the on-shell action. 
In the present work, we do not need to worry about these complications, as these counter
terms are real, and do not contribute to the spectral functions.  

The equation (\ref{OnShellGaugeinv}) is the final form for the action in terms of the 
gauge invariant variables.  Taking the imaginary part of the action
\begin{equation}
  \mbox{Im }\mathcal{S}_2(w) = \frac{1}{2i} \left(\mathcal{S}_2(w) - \mathcal{S}_2(-w)\right),
\end{equation}
and using the relations for temperature and entropy
\begin{eqnarray}
  s &=& \frac{2 \pi g_{xx}(z_h)^{3/2}}{\kappa} \\
  T &=& -\frac{f'(z_h)}{4 \pi},
\end{eqnarray}
we arrive at the form quoted in the text, Eq. (\ref{ImOnShellGaugeInv}).

\section{Derivation of the equation for $\Z_{\Phi}$ for a single scalar}
\label{SingleScalar_app}
In this section we provide the detailed steps which allow one to derive 
(\ref{Zeqnsingle}) from (\ref{multiZphi}) in the case of 
a single scalar field.  
Let us write equation (\ref{multiZphi}) as
\begin{equation}
  \e1_{Z} = 0
\end{equation}
with
\begin{eqnarray}
  \nonumber \\
  \e1_{Z} &\equiv& 
  	\frac{1}{\rootg}  \partial_z \left[\rootg g^{zz} \Z_{\Phi}'\right] 
	-w^2 g^{tt} \Z_{\Phi}  \nonumber \\
	&-&\Z_{\Phi}\left\{ \frac{\partial^2 V}{\partial \Phi^2}
	+ \frac{2}{3 \rootg }  \partial_z  \left[\rootg g^{zz} \Phi' k \right] \right\}
	.
\end{eqnarray}
We now add the following combination of the background equations (\ref{E01def}) - (\ref{E04def}) to get
\begin{eqnarray}
  \e1_Z &+& \frac{\Z_{\Phi}}{\DL[g_{xx}]} \left\{ 
  \frac{1}{k} \partial_z (\E0_{\Phi})
  +2 \DL[g_{xx}] \E0_3 \right.
  \nonumber \\
  &+& \left. \frac{1}{3} \frac{g_{zz}}{\left(k \rootg \right)^2} \partial_z \left[\left(k \rootg g^{zz}\right)^2 \E0_2 \right]
  \right\} = 0.
\end{eqnarray}
Carrying out the algebra, we find that this reduces to
\begin{eqnarray}
  &&\frac{1}{\rootg} \partial_z \left[ \rootg g^{zz} \Z_{\Phi}'  \right] 
  \nonumber \\
  &-& \Z_{\Phi} \left\{ w^2 g^{tt} 
  + \frac{1}{f k \rootg} \partial_z \left[ \rootg g^{zz} f k' \right] \right\} =0,
\end{eqnarray}
which is the equation presented in the text (\ref{Zeqnsingle}).  

\section{Chamblin-Reall spectral function - comparison with Gubser \lowercase{\emph{et al}}.}
\label{Gubser_app}
In this section, we compute the spectral density $\pbulk$ in the
Chamblin-Reall model using the method of Gubser et
al. \cite{Gubser:2008sz}.  The results are consistent with those given
in Sec. \ref{spectral_Euclidean_subsec}, though our methods are more
generally applicable to theories with multiple scalar fields.  The
notations and conventions of this section are not the same as those
given in the rest of the present work.  Instead, the notation here is
the same as that in \cite{Gubser:2008sz}.  At times we have
inserted some dimensionful factors for clarity.  

The potential is written as
\begin{equation}
	V(\Phi) = V_0 e^{\gamma \Phi}
	\label{GubserPotential}
\end{equation}
with $0 \leq \gamma \leq \sqrt{2/3}$.  
The background is written in coordinates
where the scalar field is a scaled version of the radial coordinate (which we
will call $r$ throughout this section):
\begin{eqnarray}
  ds^2 &=& e^{2A(r)} \left[ -h(r) dt^2 + d\vec{x}^2 \right] + e^{2B(r)} \frac{dr^2}{h(r)}\\
  A(r) &=& -\frac{\sqrt{-V_0}}{3\gamma} r\\
  B(r) &=& \frac{1}{2} \left( \log \left[\frac{8-3\gamma^2}{6\gamma^2} \right]-r \gamma \sqrt{-V_0} \right) \\
  h(r) &=& 1- \exp \left\{\frac{\sqrt{-V_0}(8-3 \gamma^2)}{6 \gamma} (r-r_h) \right\}\\
  \Phi(r) &=& r \sqrt{-V_0}.
\end{eqnarray}
There is a horizon at $r = r_h$.  Primes always denote derivatives
with respect to the radial coordinate.  The spectral densities are
given by
\begin{eqnarray}
  \pshear &=& \frac{e^{4A -B} h}{32 \pi i G_5 \sqrt{-V_0}} \left( h_{12}^* h_{12}' - h_{12}^{*'} h_{12} \right) \\
  \pbulk &=& \frac{e^{4A -B} h \sqrt{-V_0}}{32 \pi i G_5 A'^2} \left( h_{11}^* h_{11}' - h_{11}^{*'} h_{11} \right).
\end{eqnarray}
The above quantities are independent of $r$; the right hand side can
be evaluated at any value of $r$.  The perturbations $h_{12}$
and $h_{11}$ are found by solving the linearized Einstein equations
with the incoming wave boundary condition.  These equations are:
\begin{eqnarray}
  h_{12}''(r) &+& h_{12}'(r) \left[4 A'(r) -B'(r) +\frac{h'(r)}{h(r)}\right] \nonumber \\
  &+& \left(\frac{w e^{B(r)-A(r)}}{h(r)}\right)^2 h_{12}(r) = 0,
\end{eqnarray}
and,
\begin{eqnarray}
  \nonumber \\
  &&h_{11}''(r)=\left[\frac{V_0}{3A'(r)}-4A'(r)+3B'(r)-\frac{h'(r)}{h(r)}\right]h_{11}'(r) \nonumber \\
  &-&\left[\left(\frac{w e^{B(r)-A(r)}}{h(r)}\right)^2+\frac{V_0 h'(r)}{6h(r)A'(r)}+\frac{h'(r)B'(r)}{h(r)}\right]h_{11}(r)
  \nonumber \\
\end{eqnarray}
These equations hold for any general single scalar gravity dual.  The
key observation is that for the Chamblin-Reall background,
\begin{equation}
  \frac{V_0}{3A'(r)} + 2 B'(r) = 0.
\end{equation}
With the use of this equality, one sees that the equation for $h_{12}$
becomes identical to that of $h_{11}$.  Because the differential
equations are the same, and the boundary conditions are the same, we
conclude that \emph{for this background only}, $h_{11} = h_{12}$, and
thus, we again see that the bulk and shear spectral densities are simply related
\begin{equation}
  \frac{\pbulk(w)}{\pshear(w)} = -\frac{V_0}{A'(r)^2} = 9 \gamma^2.
\end{equation}
By comparing the definition of the potentials (\ref{potential2}), (\ref{GubserPotential}) we see that
\begin{equation}
  \gamma^2 = \frac{4 \delta}{3}.
\end{equation}
Again we find
\begin{equation}
  \frac{\pbulk(w)}{\pshear(w)} = 12 \delta.
\end{equation}
This is in agreement with the result given in the text (\ref{spectralratio}).

\bibliography{c:/Users/Todd/Documents/Physics/AdS_QCD/Drafts/Bibliography_Files/AdSCFT}

\begin{thebibliography}{53}%
\makeatletter
\providecommand \@ifxundefined [1]{%
 \@ifx{#1\undefined}
}%
\providecommand \@ifnum [1]{%
 \ifnum #1\expandafter \@firstoftwo
 \else \expandafter \@secondoftwo
 \fi
}%
\providecommand \@ifx [1]{%
 \ifx #1\expandafter \@firstoftwo
 \else \expandafter \@secondoftwo
 \fi
}%
\providecommand \natexlab [1]{#1}%
\providecommand \enquote  [1]{``#1''}%
\providecommand \bibnamefont  [1]{#1}%
\providecommand \bibfnamefont [1]{#1}%
\providecommand \citenamefont [1]{#1}%
\providecommand \href@noop [0]{\@secondoftwo}%
\providecommand \href [0]{\begingroup \@sanitize@url \@href}%
\providecommand \@href[1]{\@@startlink{#1}\@@href}%
\providecommand \@@href[1]{\endgroup#1\@@endlink}%
\providecommand \@sanitize@url [0]{\catcode `\\12\catcode `\$12\catcode
  `\&12\catcode `\#12\catcode `\^12\catcode `\_12\catcode `\%12\relax}%
\providecommand \@@startlink[1]{}%
\providecommand \@@endlink[0]{}%
\providecommand \url  [0]{\begingroup\@sanitize@url \@url }%
\providecommand \@url [1]{\endgroup\@href {#1}{\urlprefix }}%
\providecommand \urlprefix  [0]{URL }%
\providecommand \Eprint [0]{\href }%
\providecommand \doibase [0]{http://dx.doi.org/}%
\providecommand \selectlanguage [0]{\@gobble}%
\providecommand \bibinfo  [0]{\@secondoftwo}%
\providecommand \bibfield  [0]{\@secondoftwo}%
\providecommand \translation [1]{[#1]}%
\providecommand \BibitemOpen [0]{}%
\providecommand \bibitemStop [0]{}%
\providecommand \bibitemNoStop [0]{.\EOS\space}%
\providecommand \EOS [0]{\spacefactor3000\relax}%
\providecommand \BibitemShut  [1]{\csname bibitem#1\endcsname}%
\let\auto@bib@innerbib\@empty
\bibitem [{\citenamefont {Maldacena}(1998)}]{Maldacena:1997re}%
  \BibitemOpen
  \bibfield  {author} {\bibinfo {author} {\bibfnamefont {J.~M.}\ \bibnamefont
  {Maldacena}},\ }\href@noop {} {\bibfield  {journal} {\bibinfo  {journal}
  {Adv. Theor. Math. Phys.}\ }\textbf {\bibinfo {volume} {2}},\ \bibinfo
  {pages} {231} (\bibinfo {year} {1998})},\ \Eprint
  {http://arxiv.org/abs/hep-th/9711200} {arXiv:hep-th/9711200} \BibitemShut
  {NoStop}%
\bibitem [{\citenamefont {Witten}(1998)}]{Witten:1998qj}%
  \BibitemOpen
  \bibfield  {author} {\bibinfo {author} {\bibfnamefont {E.}~\bibnamefont
  {Witten}},\ }\href@noop {} {\bibfield  {journal} {\bibinfo  {journal} {Adv.
  Theor. Math. Phys.}\ }\textbf {\bibinfo {volume} {2}},\ \bibinfo {pages}
  {253} (\bibinfo {year} {1998})},\ \Eprint
  {http://arxiv.org/abs/hep-th/9802150} {arXiv:hep-th/9802150} \BibitemShut
  {NoStop}%
\bibitem [{\citenamefont {Gubser}\ \emph {et~al.}(1998)\citenamefont {Gubser},
  \citenamefont {Klebanov},\ and\ \citenamefont {Polyakov}}]{Gubser:1998bc}%
  \BibitemOpen
  \bibfield  {author} {\bibinfo {author} {\bibfnamefont {S.~S.}\ \bibnamefont
  {Gubser}}, \bibinfo {author} {\bibfnamefont {I.~R.}\ \bibnamefont
  {Klebanov}}, \ and\ \bibinfo {author} {\bibfnamefont {A.~M.}\ \bibnamefont
  {Polyakov}},\ }\href {\doibase 10.1016/S0370-2693(98)00377-3} {\bibfield
  {journal} {\bibinfo  {journal} {Phys. Lett.}\ }\textbf {\bibinfo {volume}
  {B428}},\ \bibinfo {pages} {105} (\bibinfo {year} {1998})},\ \Eprint
  {http://arxiv.org/abs/hep-th/9802109} {arXiv:hep-th/9802109} \BibitemShut
  {NoStop}%
\bibitem [{\citenamefont {Klebanov}\ and\ \citenamefont
  {Witten}(1999)}]{Klebanov:1999tb}%
  \BibitemOpen
  \bibfield  {author} {\bibinfo {author} {\bibfnamefont {I.~R.}\ \bibnamefont
  {Klebanov}}\ and\ \bibinfo {author} {\bibfnamefont {E.}~\bibnamefont
  {Witten}},\ }\href {\doibase 10.1016/S0550-3213(99)00387-9} {\bibfield
  {journal} {\bibinfo  {journal} {Nucl. Phys.}\ }\textbf {\bibinfo {volume}
  {B556}},\ \bibinfo {pages} {89} (\bibinfo {year} {1999})},\ \Eprint
  {http://arxiv.org/abs/hep-th/9905104} {arXiv:hep-th/9905104} \BibitemShut
  {NoStop}%
\bibitem [{\citenamefont {Aharony}\ \emph {et~al.}(2000)\citenamefont
  {Aharony}, \citenamefont {Gubser}, \citenamefont {Maldacena}, \citenamefont
  {Ooguri},\ and\ \citenamefont {Oz}}]{Aharony:1999ti}%
  \BibitemOpen
  \bibfield  {author} {\bibinfo {author} {\bibfnamefont {O.}~\bibnamefont
  {Aharony}}, \bibinfo {author} {\bibfnamefont {S.~S.}\ \bibnamefont {Gubser}},
  \bibinfo {author} {\bibfnamefont {J.~M.}\ \bibnamefont {Maldacena}}, \bibinfo
  {author} {\bibfnamefont {H.}~\bibnamefont {Ooguri}}, \ and\ \bibinfo {author}
  {\bibfnamefont {Y.}~\bibnamefont {Oz}},\ }\href {\doibase
  10.1016/S0370-1573(99)00083-6} {\bibfield  {journal} {\bibinfo  {journal}
  {Phys. Rept.}\ }\textbf {\bibinfo {volume} {323}},\ \bibinfo {pages} {183}
  (\bibinfo {year} {2000})},\ \Eprint {http://arxiv.org/abs/hep-th/9905111}
  {arXiv:hep-th/9905111} \BibitemShut {NoStop}%
\bibitem [{\citenamefont {Son}\ and\ \citenamefont
  {Starinets}(2007)}]{Son:2007vk}%
  \BibitemOpen
  \bibfield  {author} {\bibinfo {author} {\bibfnamefont {D.~T.}\ \bibnamefont
  {Son}}\ and\ \bibinfo {author} {\bibfnamefont {A.~O.}\ \bibnamefont
  {Starinets}},\ }\href {\doibase 10.1146/annurev.nucl.57.090506.123120}
  {\bibfield  {journal} {\bibinfo  {journal} {Ann. Rev. Nucl. Part. Sci.}\
  }\textbf {\bibinfo {volume} {57}},\ \bibinfo {pages} {95} (\bibinfo {year}
  {2007})},\ \Eprint {http://arxiv.org/abs/0704.0240} {arXiv:0704.0240
  [hep-th]} \BibitemShut {NoStop}%
\bibitem [{\citenamefont {Erdmenger}\ \emph {et~al.}(2008)\citenamefont
  {Erdmenger}, \citenamefont {Evans}, \citenamefont {Kirsch},\ and\
  \citenamefont {Threlfall}}]{Erdmenger:2007cm}%
  \BibitemOpen
  \bibfield  {author} {\bibinfo {author} {\bibfnamefont {J.}~\bibnamefont
  {Erdmenger}}, \bibinfo {author} {\bibfnamefont {N.}~\bibnamefont {Evans}},
  \bibinfo {author} {\bibfnamefont {I.}~\bibnamefont {Kirsch}}, \ and\ \bibinfo
  {author} {\bibfnamefont {E.}~\bibnamefont {Threlfall}},\ }\href {\doibase
  10.1140/epja/i2007-10540-1} {\bibfield  {journal} {\bibinfo  {journal} {Eur.
  Phys. J.}\ }\textbf {\bibinfo {volume} {A35}},\ \bibinfo {pages} {81}
  (\bibinfo {year} {2008})},\ \Eprint {http://arxiv.org/abs/0711.4467}
  {arXiv:0711.4467 [hep-th]} \BibitemShut {NoStop}%
\bibitem [{\citenamefont {Myers}\ and\ \citenamefont
  {Vazquez}(2008)}]{Myers:2008fv}%
  \BibitemOpen
  \bibfield  {author} {\bibinfo {author} {\bibfnamefont {R.~C.}\ \bibnamefont
  {Myers}}\ and\ \bibinfo {author} {\bibfnamefont {S.~E.}\ \bibnamefont
  {Vazquez}},\ }\href {\doibase 10.1088/0264-9381/25/11/114008} {\bibfield
  {journal} {\bibinfo  {journal} {Class. Quant. Grav.}\ }\textbf {\bibinfo
  {volume} {25}},\ \bibinfo {pages} {114008} (\bibinfo {year} {2008})},\
  \Eprint {http://arxiv.org/abs/0804.2423} {arXiv:0804.2423 [hep-th]}
  \BibitemShut {NoStop}%
\bibitem [{\citenamefont {Gubser}\ and\ \citenamefont
  {Karch}(2009)}]{Gubser:2009md}%
  \BibitemOpen
  \bibfield  {author} {\bibinfo {author} {\bibfnamefont {S.~S.}\ \bibnamefont
  {Gubser}}\ and\ \bibinfo {author} {\bibfnamefont {A.}~\bibnamefont {Karch}},\
  }\href {\doibase 10.1146/annurev.nucl.010909.083602} {\bibfield  {journal}
  {\bibinfo  {journal} {Ann. Rev. Nucl. Part. Sci.}\ }\textbf {\bibinfo
  {volume} {59}},\ \bibinfo {pages} {145} (\bibinfo {year} {2009})},\ \Eprint
  {http://arxiv.org/abs/0901.0935} {arXiv:0901.0935 [hep-th]} \BibitemShut
  {NoStop}%
\bibitem [{\citenamefont {Klebanov}\ and\ \citenamefont
  {Tseytlin}(2000)}]{Klebanov:2000nc}%
  \BibitemOpen
  \bibfield  {author} {\bibinfo {author} {\bibfnamefont {I.~R.}\ \bibnamefont
  {Klebanov}}\ and\ \bibinfo {author} {\bibfnamefont {A.~A.}\ \bibnamefont
  {Tseytlin}},\ }\href {\doibase 10.1016/S0550-3213(00)00206-6} {\bibfield
  {journal} {\bibinfo  {journal} {Nucl. Phys.}\ }\textbf {\bibinfo {volume}
  {B578}},\ \bibinfo {pages} {123} (\bibinfo {year} {2000})},\ \Eprint
  {http://arxiv.org/abs/hep-th/0002159} {arXiv:hep-th/0002159} \BibitemShut
  {NoStop}%
\bibitem [{\citenamefont {Klebanov}\ and\ \citenamefont
  {Strassler}(2000)}]{Klebanov:2000hb}%
  \BibitemOpen
  \bibfield  {author} {\bibinfo {author} {\bibfnamefont {I.~R.}\ \bibnamefont
  {Klebanov}}\ and\ \bibinfo {author} {\bibfnamefont {M.~J.}\ \bibnamefont
  {Strassler}},\ }\href@noop {} {\bibfield  {journal} {\bibinfo  {journal}
  {JHEP}\ }\textbf {\bibinfo {volume} {08}},\ \bibinfo {pages} {052} (\bibinfo
  {year} {2000})},\ \Eprint {http://arxiv.org/abs/hep-th/0007191}
  {arXiv:hep-th/0007191} \BibitemShut {NoStop}%
\bibitem [{\citenamefont {Pilch}\ and\ \citenamefont
  {Warner}(2001)}]{Pilch:2000ue}%
  \BibitemOpen
  \bibfield  {author} {\bibinfo {author} {\bibfnamefont {K.}~\bibnamefont
  {Pilch}}\ and\ \bibinfo {author} {\bibfnamefont {N.~P.}\ \bibnamefont
  {Warner}},\ }\href {\doibase 10.1016/S0550-3213(00)00656-8} {\bibfield
  {journal} {\bibinfo  {journal} {Nucl. Phys.}\ }\textbf {\bibinfo {volume}
  {B594}},\ \bibinfo {pages} {209} (\bibinfo {year} {2001})},\ \Eprint
  {http://arxiv.org/abs/hep-th/0004063} {arXiv:hep-th/0004063} \BibitemShut
  {NoStop}%
\bibitem [{\citenamefont {Mia}\ \emph {et~al.}(2010)\citenamefont {Mia},
  \citenamefont {Dasgupta}, \citenamefont {Gale},\ and\ \citenamefont
  {Jeon}}]{Mia:2009wj}%
  \BibitemOpen
  \bibfield  {author} {\bibinfo {author} {\bibfnamefont {M.}~\bibnamefont
  {Mia}}, \bibinfo {author} {\bibfnamefont {K.}~\bibnamefont {Dasgupta}},
  \bibinfo {author} {\bibfnamefont {C.}~\bibnamefont {Gale}}, \ and\ \bibinfo
  {author} {\bibfnamefont {S.}~\bibnamefont {Jeon}},\ }\href {\doibase
  10.1016/j.nuclphysb.2010.06.014} {\bibfield  {journal} {\bibinfo  {journal}
  {Nucl. Phys.}\ }\textbf {\bibinfo {volume} {B839}},\ \bibinfo {pages} {187}
  (\bibinfo {year} {2010})},\ \Eprint {http://arxiv.org/abs/0902.1540}
  {arXiv:0902.1540 [hep-th]} \BibitemShut {NoStop}%
\bibitem [{\citenamefont {Gubser}\ \emph
  {et~al.}(2008{\natexlab{a}})\citenamefont {Gubser}, \citenamefont {Pufu},\
  and\ \citenamefont {Rocha}}]{Gubser:2008sz}%
  \BibitemOpen
  \bibfield  {author} {\bibinfo {author} {\bibfnamefont {S.~S.}\ \bibnamefont
  {Gubser}}, \bibinfo {author} {\bibfnamefont {S.~S.}\ \bibnamefont {Pufu}}, \
  and\ \bibinfo {author} {\bibfnamefont {F.~D.}\ \bibnamefont {Rocha}},\ }\href
  {\doibase 10.1088/1126-6708/2008/08/085} {\bibfield  {journal} {\bibinfo
  {journal} {JHEP}\ }\textbf {\bibinfo {volume} {08}},\ \bibinfo {pages} {085}
  (\bibinfo {year} {2008}{\natexlab{a}})},\ \Eprint
  {http://arxiv.org/abs/0806.0407} {arXiv:0806.0407 [hep-th]} \BibitemShut
  {NoStop}%
\bibitem [{\citenamefont {Gubser}\ and\ \citenamefont
  {Nellore}(2008)}]{Gubser:2008ny}%
  \BibitemOpen
  \bibfield  {author} {\bibinfo {author} {\bibfnamefont {S.~S.}\ \bibnamefont
  {Gubser}}\ and\ \bibinfo {author} {\bibfnamefont {A.}~\bibnamefont
  {Nellore}},\ }\href {\doibase 10.1103/PhysRevD.78.086007} {\bibfield
  {journal} {\bibinfo  {journal} {Phys. Rev.}\ }\textbf {\bibinfo {volume}
  {D78}},\ \bibinfo {pages} {086007} (\bibinfo {year} {2008})},\ \Eprint
  {http://arxiv.org/abs/0804.0434} {arXiv:0804.0434 [hep-th]} \BibitemShut
  {NoStop}%
\bibitem [{\citenamefont {Gursoy}\ and\ \citenamefont
  {Kiritsis}(2008)}]{Gursoy:2007cb}%
  \BibitemOpen
  \bibfield  {author} {\bibinfo {author} {\bibfnamefont {U.}~\bibnamefont
  {Gursoy}}\ and\ \bibinfo {author} {\bibfnamefont {E.}~\bibnamefont
  {Kiritsis}},\ }\href {\doibase 10.1088/1126-6708/2008/02/032} {\bibfield
  {journal} {\bibinfo  {journal} {JHEP}\ }\textbf {\bibinfo {volume} {02}},\
  \bibinfo {pages} {032} (\bibinfo {year} {2008})},\ \Eprint
  {http://arxiv.org/abs/0707.1324} {arXiv:0707.1324 [hep-th]} \BibitemShut
  {NoStop}%
\bibitem [{\citenamefont {Gursoy}\ \emph {et~al.}(2008)\citenamefont {Gursoy},
  \citenamefont {Kiritsis},\ and\ \citenamefont {Nitti}}]{Gursoy:2007er}%
  \BibitemOpen
  \bibfield  {author} {\bibinfo {author} {\bibfnamefont {U.}~\bibnamefont
  {Gursoy}}, \bibinfo {author} {\bibfnamefont {E.}~\bibnamefont {Kiritsis}}, \
  and\ \bibinfo {author} {\bibfnamefont {F.}~\bibnamefont {Nitti}},\ }\href
  {\doibase 10.1088/1126-6708/2008/02/019} {\bibfield  {journal} {\bibinfo
  {journal} {JHEP}\ }\textbf {\bibinfo {volume} {02}},\ \bibinfo {pages} {019}
  (\bibinfo {year} {2008})},\ \Eprint {http://arxiv.org/abs/0707.1349}
  {arXiv:0707.1349 [hep-th]} \BibitemShut {NoStop}%
\bibitem [{\citenamefont {Gursoy}\ \emph
  {et~al.}(2009{\natexlab{a}})\citenamefont {Gursoy}, \citenamefont {Kiritsis},
  \citenamefont {Mazzanti},\ and\ \citenamefont {Nitti}}]{Gursoy:2008za}%
  \BibitemOpen
  \bibfield  {author} {\bibinfo {author} {\bibfnamefont {U.}~\bibnamefont
  {Gursoy}}, \bibinfo {author} {\bibfnamefont {E.}~\bibnamefont {Kiritsis}},
  \bibinfo {author} {\bibfnamefont {L.}~\bibnamefont {Mazzanti}}, \ and\
  \bibinfo {author} {\bibfnamefont {F.}~\bibnamefont {Nitti}},\ }\href
  {\doibase 10.1088/1126-6708/2009/05/033} {\bibfield  {journal} {\bibinfo
  {journal} {JHEP}\ }\textbf {\bibinfo {volume} {05}},\ \bibinfo {pages} {033}
  (\bibinfo {year} {2009}{\natexlab{a}})},\ \Eprint
  {http://arxiv.org/abs/0812.0792} {arXiv:0812.0792 [hep-th]} \BibitemShut
  {NoStop}%
\bibitem [{\citenamefont {Gursoy}\ \emph
  {et~al.}(2009{\natexlab{b}})\citenamefont {Gursoy}, \citenamefont {Kiritsis},
  \citenamefont {Michalogiorgakis},\ and\ \citenamefont
  {Nitti}}]{Gursoy:2009kk}%
  \BibitemOpen
  \bibfield  {author} {\bibinfo {author} {\bibfnamefont {U.}~\bibnamefont
  {Gursoy}}, \bibinfo {author} {\bibfnamefont {E.}~\bibnamefont {Kiritsis}},
  \bibinfo {author} {\bibfnamefont {G.}~\bibnamefont {Michalogiorgakis}}, \
  and\ \bibinfo {author} {\bibfnamefont {F.}~\bibnamefont {Nitti}},\ }\href
  {\doibase 10.1088/1126-6708/2009/12/056} {\bibfield  {journal} {\bibinfo
  {journal} {JHEP}\ }\textbf {\bibinfo {volume} {12}},\ \bibinfo {pages} {056}
  (\bibinfo {year} {2009}{\natexlab{b}})},\ \Eprint
  {http://arxiv.org/abs/0906.1890} {arXiv:0906.1890 [hep-ph]} \BibitemShut
  {NoStop}%
\bibitem [{\citenamefont {Springer}(2009{\natexlab{a}})}]{Springer:2008js}%
  \BibitemOpen
  \bibfield  {author} {\bibinfo {author} {\bibfnamefont {T.}~\bibnamefont
  {Springer}},\ }\href {\doibase 10.1103/PhysRevD.79.046003} {\bibfield
  {journal} {\bibinfo  {journal} {Phys. Rev.}\ }\textbf {\bibinfo {volume}
  {D79}},\ \bibinfo {pages} {046003} (\bibinfo {year} {2009}{\natexlab{a}})},\
  \Eprint {http://arxiv.org/abs/0810.4354} {arXiv:0810.4354 [hep-th]}
  \BibitemShut {NoStop}%
\bibitem [{\citenamefont {Springer}(2009{\natexlab{b}})}]{Springer:2009wj}%
  \BibitemOpen
  \bibfield  {author} {\bibinfo {author} {\bibfnamefont {T.}~\bibnamefont
  {Springer}},\ }\href {\doibase 10.1103/PhysRevD.79.086003} {\bibfield
  {journal} {\bibinfo  {journal} {Phys. Rev.}\ }\textbf {\bibinfo {volume}
  {D79}},\ \bibinfo {pages} {086003} (\bibinfo {year} {2009}{\natexlab{b}})},\
  \Eprint {http://arxiv.org/abs/0902.2566} {arXiv:0902.2566 [hep-th]}
  \BibitemShut {NoStop}%
\bibitem [{\citenamefont {Batell}\ and\ \citenamefont
  {Gherghetta}(2008)}]{Batell:2008zm}%
  \BibitemOpen
  \bibfield  {author} {\bibinfo {author} {\bibfnamefont {B.}~\bibnamefont
  {Batell}}\ and\ \bibinfo {author} {\bibfnamefont {T.}~\bibnamefont
  {Gherghetta}},\ }\href {\doibase 10.1103/PhysRevD.78.026002} {\bibfield
  {journal} {\bibinfo  {journal} {Phys. Rev.}\ }\textbf {\bibinfo {volume}
  {D78}},\ \bibinfo {pages} {026002} (\bibinfo {year} {2008})},\ \Eprint
  {http://arxiv.org/abs/0801.4383} {arXiv:0801.4383 [hep-ph]} \BibitemShut
  {NoStop}%
\bibitem [{\citenamefont {de~Paula}\ \emph {et~al.}(2009)\citenamefont
  {de~Paula}, \citenamefont {Frederico}, \citenamefont {Forkel},\ and\
  \citenamefont {Beyer}}]{dePaula:2008fp}%
  \BibitemOpen
  \bibfield  {author} {\bibinfo {author} {\bibfnamefont {W.}~\bibnamefont
  {de~Paula}}, \bibinfo {author} {\bibfnamefont {T.}~\bibnamefont {Frederico}},
  \bibinfo {author} {\bibfnamefont {H.}~\bibnamefont {Forkel}}, \ and\ \bibinfo
  {author} {\bibfnamefont {M.}~\bibnamefont {Beyer}},\ }\href {\doibase
  10.1103/PhysRevD.79.075019} {\bibfield  {journal} {\bibinfo  {journal} {Phys.
  Rev.}\ }\textbf {\bibinfo {volume} {D79}},\ \bibinfo {pages} {075019}
  (\bibinfo {year} {2009})},\ \Eprint {http://arxiv.org/abs/0806.3830}
  {arXiv:0806.3830 [hep-ph]} \BibitemShut {NoStop}%
\bibitem [{\citenamefont {Alanen}\ \emph {et~al.}(2009)\citenamefont {Alanen},
  \citenamefont {Kajantie},\ and\ \citenamefont {Suur-Uski}}]{Alanen:2009xs}%
  \BibitemOpen
  \bibfield  {author} {\bibinfo {author} {\bibfnamefont {J.}~\bibnamefont
  {Alanen}}, \bibinfo {author} {\bibfnamefont {K.}~\bibnamefont {Kajantie}}, \
  and\ \bibinfo {author} {\bibfnamefont {V.}~\bibnamefont {Suur-Uski}},\ }\href
  {\doibase 10.1103/PhysRevD.80.126008} {\bibfield  {journal} {\bibinfo
  {journal} {Phys. Rev.}\ }\textbf {\bibinfo {volume} {D80}},\ \bibinfo {pages}
  {126008} (\bibinfo {year} {2009})},\ \Eprint {http://arxiv.org/abs/0911.2114}
  {arXiv:0911.2114 [hep-ph]} \BibitemShut {NoStop}%
\bibitem [{\citenamefont {Erlich}\ \emph {et~al.}(2005)\citenamefont {Erlich},
  \citenamefont {Katz}, \citenamefont {Son},\ and\ \citenamefont
  {Stephanov}}]{Erlich:2005qh}%
  \BibitemOpen
  \bibfield  {author} {\bibinfo {author} {\bibfnamefont {J.}~\bibnamefont
  {Erlich}}, \bibinfo {author} {\bibfnamefont {E.}~\bibnamefont {Katz}},
  \bibinfo {author} {\bibfnamefont {D.~T.}\ \bibnamefont {Son}}, \ and\
  \bibinfo {author} {\bibfnamefont {M.~A.}\ \bibnamefont {Stephanov}},\ }\href
  {\doibase 10.1103/PhysRevLett.95.261602} {\bibfield  {journal} {\bibinfo
  {journal} {Phys. Rev. Lett.}\ }\textbf {\bibinfo {volume} {95}},\ \bibinfo
  {pages} {261602} (\bibinfo {year} {2005})},\ \Eprint
  {http://arxiv.org/abs/hep-ph/0501128} {arXiv:hep-ph/0501128} \BibitemShut
  {NoStop}%
\bibitem [{\citenamefont {Kapusta}\ and\ \citenamefont
  {Springer}(2010)}]{Kapusta:2010mf}%
  \BibitemOpen
  \bibfield  {author} {\bibinfo {author} {\bibfnamefont {J.~I.}\ \bibnamefont
  {Kapusta}}\ and\ \bibinfo {author} {\bibfnamefont {T.}~\bibnamefont
  {Springer}},\ }\href {\doibase 10.1103/PhysRevD.81.086009} {\bibfield
  {journal} {\bibinfo  {journal} {Phys. Rev.}\ }\textbf {\bibinfo {volume}
  {D81}},\ \bibinfo {pages} {086009} (\bibinfo {year} {2010})},\ \Eprint
  {http://arxiv.org/abs/1001.4799} {arXiv:1001.4799 [hep-ph]} \BibitemShut
  {NoStop}%
\bibitem [{\citenamefont {Benincasa}\ \emph {et~al.}(2006)\citenamefont
  {Benincasa}, \citenamefont {Buchel},\ and\ \citenamefont
  {Starinets}}]{Benincasa:2005iv}%
  \BibitemOpen
  \bibfield  {author} {\bibinfo {author} {\bibfnamefont {P.}~\bibnamefont
  {Benincasa}}, \bibinfo {author} {\bibfnamefont {A.}~\bibnamefont {Buchel}}, \
  and\ \bibinfo {author} {\bibfnamefont {A.~O.}\ \bibnamefont {Starinets}},\
  }\href {\doibase 10.1016/j.nuclphysb.2005.11.005} {\bibfield  {journal}
  {\bibinfo  {journal} {Nucl. Phys.}\ }\textbf {\bibinfo {volume} {B733}},\
  \bibinfo {pages} {160} (\bibinfo {year} {2006})},\ \Eprint
  {http://arxiv.org/abs/hep-th/0507026} {arXiv:hep-th/0507026} \BibitemShut
  {NoStop}%
\bibitem [{\citenamefont {Aharony}\ \emph {et~al.}(2005)\citenamefont
  {Aharony}, \citenamefont {Buchel},\ and\ \citenamefont
  {Yarom}}]{Aharony:2005zr}%
  \BibitemOpen
  \bibfield  {author} {\bibinfo {author} {\bibfnamefont {O.}~\bibnamefont
  {Aharony}}, \bibinfo {author} {\bibfnamefont {A.}~\bibnamefont {Buchel}}, \
  and\ \bibinfo {author} {\bibfnamefont {A.}~\bibnamefont {Yarom}},\ }\href
  {\doibase 10.1103/PhysRevD.72.066003} {\bibfield  {journal} {\bibinfo
  {journal} {Phys. Rev.}\ }\textbf {\bibinfo {volume} {D72}},\ \bibinfo {pages}
  {066003} (\bibinfo {year} {2005})},\ \Eprint
  {http://arxiv.org/abs/hep-th/0506002} {arXiv:hep-th/0506002} \BibitemShut
  {NoStop}%
\bibitem [{\citenamefont {Meyer}(2007)}]{Meyer:2007ic}%
  \BibitemOpen
  \bibfield  {author} {\bibinfo {author} {\bibfnamefont {H.~B.}\ \bibnamefont
  {Meyer}},\ }\href {\doibase 10.1103/PhysRevD.76.101701} {\bibfield  {journal}
  {\bibinfo  {journal} {Phys. Rev.}\ }\textbf {\bibinfo {volume} {D76}},\
  \bibinfo {pages} {101701} (\bibinfo {year} {2007})},\ \Eprint
  {http://arxiv.org/abs/0704.1801} {arXiv:0704.1801 [hep-lat]} \BibitemShut
  {NoStop}%
\bibitem [{\citenamefont {Meyer}(2010)}]{Meyer:2010ii}%
  \BibitemOpen
  \bibfield  {author} {\bibinfo {author} {\bibfnamefont {H.~B.}\ \bibnamefont
  {Meyer}},\ }\href {\doibase 10.1007/JHEP04(2010)099} {\bibfield  {journal}
  {\bibinfo  {journal} {JHEP}\ }\textbf {\bibinfo {volume} {04}},\ \bibinfo
  {pages} {099} (\bibinfo {year} {2010})},\ \Eprint
  {http://arxiv.org/abs/1002.3343} {arXiv:1002.3343 [hep-lat]} \BibitemShut
  {NoStop}%
\bibitem [{\citenamefont {Kharzeev}\ and\ \citenamefont
  {Tuchin}(2008)}]{Kharzeev:2007wb}%
  \BibitemOpen
  \bibfield  {author} {\bibinfo {author} {\bibfnamefont {D.}~\bibnamefont
  {Kharzeev}}\ and\ \bibinfo {author} {\bibfnamefont {K.}~\bibnamefont
  {Tuchin}},\ }\href {\doibase 10.1088/1126-6708/2008/09/093} {\bibfield
  {journal} {\bibinfo  {journal} {JHEP}\ }\textbf {\bibinfo {volume} {09}},\
  \bibinfo {pages} {093} (\bibinfo {year} {2008})},\ \Eprint
  {http://arxiv.org/abs/0705.4280} {arXiv:0705.4280 [hep-ph]} \BibitemShut
  {NoStop}%
\bibitem [{\citenamefont {Moore}\ and\ \citenamefont
  {Saremi}(2008)}]{Moore:2008ws}%
  \BibitemOpen
  \bibfield  {author} {\bibinfo {author} {\bibfnamefont {G.~D.}\ \bibnamefont
  {Moore}}\ and\ \bibinfo {author} {\bibfnamefont {O.}~\bibnamefont {Saremi}},\
  }\href {\doibase 10.1088/1126-6708/2008/09/015} {\bibfield  {journal}
  {\bibinfo  {journal} {JHEP}\ }\textbf {\bibinfo {volume} {09}},\ \bibinfo
  {pages} {015} (\bibinfo {year} {2008})},\ \Eprint
  {http://arxiv.org/abs/0805.4201} {arXiv:0805.4201 [hep-ph]} \BibitemShut
  {NoStop}%
\bibitem [{\citenamefont {Yarom}(2010)}]{Yarom:2009mw}%
  \BibitemOpen
  \bibfield  {author} {\bibinfo {author} {\bibfnamefont {A.}~\bibnamefont
  {Yarom}},\ }\href {\doibase 10.1007/JHEP04(2010)024} {\bibfield  {journal}
  {\bibinfo  {journal} {JHEP}\ }\textbf {\bibinfo {volume} {04}},\ \bibinfo
  {pages} {024} (\bibinfo {year} {2010})},\ \Eprint
  {http://arxiv.org/abs/0912.2100} {arXiv:0912.2100 [hep-th]} \BibitemShut
  {NoStop}%
\bibitem [{\citenamefont {Chamblin}\ and\ \citenamefont
  {Reall}(1999)}]{Chamblin:1999ya}%
  \BibitemOpen
  \bibfield  {author} {\bibinfo {author} {\bibfnamefont {H.~A.}\ \bibnamefont
  {Chamblin}}\ and\ \bibinfo {author} {\bibfnamefont {H.~S.}\ \bibnamefont
  {Reall}},\ }\href {\doibase 10.1016/S0550-3213(99)00520-9} {\bibfield
  {journal} {\bibinfo  {journal} {Nucl. Phys.}\ }\textbf {\bibinfo {volume}
  {B562}},\ \bibinfo {pages} {133} (\bibinfo {year} {1999})},\ \Eprint
  {http://arxiv.org/abs/hep-th/9903225} {arXiv:hep-th/9903225} \BibitemShut
  {NoStop}%
\bibitem [{\citenamefont {Bigazzi}\ \emph {et~al.}(2010)\citenamefont
  {Bigazzi}, \citenamefont {Cotrone},\ and\ \citenamefont
  {Tarrio}}]{Bigazzi:2009tc}%
  \BibitemOpen
  \bibfield  {author} {\bibinfo {author} {\bibfnamefont {F.}~\bibnamefont
  {Bigazzi}}, \bibinfo {author} {\bibfnamefont {A.~L.}\ \bibnamefont
  {Cotrone}}, \ and\ \bibinfo {author} {\bibfnamefont {J.}~\bibnamefont
  {Tarrio}},\ }\href {\doibase 10.1007/JHEP02(2010)083} {\bibfield  {journal}
  {\bibinfo  {journal} {JHEP}\ }\textbf {\bibinfo {volume} {02}},\ \bibinfo
  {pages} {083} (\bibinfo {year} {2010})},\ \Eprint
  {http://arxiv.org/abs/0912.3256} {arXiv:0912.3256 [hep-th]} \BibitemShut
  {NoStop}%
\bibitem [{\citenamefont {Gubser}\ \emph
  {et~al.}(2008{\natexlab{b}})\citenamefont {Gubser}, \citenamefont {Nellore},
  \citenamefont {Pufu},\ and\ \citenamefont {Rocha}}]{Gubser:2008yx}%
  \BibitemOpen
  \bibfield  {author} {\bibinfo {author} {\bibfnamefont {S.~S.}\ \bibnamefont
  {Gubser}}, \bibinfo {author} {\bibfnamefont {A.}~\bibnamefont {Nellore}},
  \bibinfo {author} {\bibfnamefont {S.~S.}\ \bibnamefont {Pufu}}, \ and\
  \bibinfo {author} {\bibfnamefont {F.~D.}\ \bibnamefont {Rocha}},\ }\href
  {\doibase 10.1103/PhysRevLett.101.131601} {\bibfield  {journal} {\bibinfo
  {journal} {Phys. Rev. Lett.}\ }\textbf {\bibinfo {volume} {101}},\ \bibinfo
  {pages} {131601} (\bibinfo {year} {2008}{\natexlab{b}})},\ \Eprint
  {http://arxiv.org/abs/0804.1950} {arXiv:0804.1950 [hep-th]} \BibitemShut
  {NoStop}%
\bibitem [{\citenamefont {Kanitscheider}\ and\ \citenamefont
  {Skenderis}(2009)}]{Kanitscheider:2009as}%
  \BibitemOpen
  \bibfield  {author} {\bibinfo {author} {\bibfnamefont {I.}~\bibnamefont
  {Kanitscheider}}\ and\ \bibinfo {author} {\bibfnamefont {K.}~\bibnamefont
  {Skenderis}},\ }\href {\doibase 10.1088/1126-6708/2009/04/062} {\bibfield
  {journal} {\bibinfo  {journal} {JHEP}\ }\textbf {\bibinfo {volume} {04}},\
  \bibinfo {pages} {062} (\bibinfo {year} {2009})},\ \Eprint
  {http://arxiv.org/abs/0901.1487} {arXiv:0901.1487 [hep-th]} \BibitemShut
  {NoStop}%
\bibitem [{\citenamefont {Romatschke}(2010)}]{Romatschke:2009kr}%
  \BibitemOpen
  \bibfield  {author} {\bibinfo {author} {\bibfnamefont {P.}~\bibnamefont
  {Romatschke}},\ }\href {\doibase 10.1088/0264-9381/27/2/025006} {\bibfield
  {journal} {\bibinfo  {journal} {Class. Quant. Grav.}\ }\textbf {\bibinfo
  {volume} {27}},\ \bibinfo {pages} {025006} (\bibinfo {year} {2010})},\
  \Eprint {http://arxiv.org/abs/0906.4787} {arXiv:0906.4787 [hep-th]}
  \BibitemShut {NoStop}%
\bibitem [{\citenamefont {Bigazzi}\ and\ \citenamefont
  {Cotrone}(2010)}]{Bigazzi:2010ku}%
  \BibitemOpen
  \bibfield  {author} {\bibinfo {author} {\bibfnamefont {F.}~\bibnamefont
  {Bigazzi}}\ and\ \bibinfo {author} {\bibfnamefont {A.~L.}\ \bibnamefont
  {Cotrone}},\ }\href {\doibase 10.1007/JHEP08(2010)128} {\bibfield  {journal}
  {\bibinfo  {journal} {JHEP}\ }\textbf {\bibinfo {volume} {1008}},\ \bibinfo
  {pages} {128} (\bibinfo {year} {2010})},\ \Eprint
  {http://arxiv.org/abs/arXiv:1006.4634} {arXiv:arXiv:1006.4634 [hep-ph]}
  \BibitemShut {NoStop}%
\bibitem [{\citenamefont {Romatschke}\ and\ \citenamefont
  {Son}(2009)}]{Romatschke:2009ng}%
  \BibitemOpen
  \bibfield  {author} {\bibinfo {author} {\bibfnamefont {P.}~\bibnamefont
  {Romatschke}}\ and\ \bibinfo {author} {\bibfnamefont {D.~T.}\ \bibnamefont
  {Son}},\ }\href {\doibase 10.1103/PhysRevD.80.065021} {\bibfield  {journal}
  {\bibinfo  {journal} {Phys. Rev.}\ }\textbf {\bibinfo {volume} {D80}},\
  \bibinfo {pages} {065021} (\bibinfo {year} {2009})},\ \Eprint
  {http://arxiv.org/abs/0903.3946} {arXiv:0903.3946 [hep-ph]} \BibitemShut
  {NoStop}%
\bibitem [{\citenamefont {Springer}\ \emph {et~al.}(2010)\citenamefont
  {Springer}, \citenamefont {Gale}, \citenamefont {Jeon},\ and\ \citenamefont
  {Lee}}]{Springer:2010mf}%
  \BibitemOpen
  \bibfield  {author} {\bibinfo {author} {\bibfnamefont {T.}~\bibnamefont
  {Springer}}, \bibinfo {author} {\bibfnamefont {C.}~\bibnamefont {Gale}},
  \bibinfo {author} {\bibfnamefont {S.}~\bibnamefont {Jeon}}, \ and\ \bibinfo
  {author} {\bibfnamefont {S.~H.}\ \bibnamefont {Lee}},\ }\href {\doibase
  10.1103/PhysRevD.82.106005} {\bibfield  {journal} {\bibinfo  {journal} {Phys.
  Rev.}\ }\textbf {\bibinfo {volume} {D82}},\ \bibinfo {pages} {106005}
  (\bibinfo {year} {2010})},\ \Eprint {http://arxiv.org/abs/1006.4667}
  {arXiv:1006.4667 [hep-th]} \BibitemShut {NoStop}%
\bibitem [{\citenamefont {Policastro}\ \emph {et~al.}(2002)\citenamefont
  {Policastro}, \citenamefont {Son},\ and\ \citenamefont
  {Starinets}}]{Policastro:2002tn}%
  \BibitemOpen
  \bibfield  {author} {\bibinfo {author} {\bibfnamefont {G.}~\bibnamefont
  {Policastro}}, \bibinfo {author} {\bibfnamefont {D.~T.}\ \bibnamefont {Son}},
  \ and\ \bibinfo {author} {\bibfnamefont {A.~O.}\ \bibnamefont {Starinets}},\
  }\href@noop {} {\bibfield  {journal} {\bibinfo  {journal} {JHEP}\ }\textbf
  {\bibinfo {volume} {12}},\ \bibinfo {pages} {054} (\bibinfo {year} {2002})},\
  \Eprint {http://arxiv.org/abs/hep-th/0210220} {arXiv:hep-th/0210220}
  \BibitemShut {NoStop}%
\bibitem [{\citenamefont {Kovtun}\ and\ \citenamefont
  {Starinets}(2005)}]{Kovtun:2005ev}%
  \BibitemOpen
  \bibfield  {author} {\bibinfo {author} {\bibfnamefont {P.~K.}\ \bibnamefont
  {Kovtun}}\ and\ \bibinfo {author} {\bibfnamefont {A.~O.}\ \bibnamefont
  {Starinets}},\ }\href {\doibase 10.1103/PhysRevD.72.086009} {\bibfield
  {journal} {\bibinfo  {journal} {Phys. Rev.}\ }\textbf {\bibinfo {volume}
  {D72}},\ \bibinfo {pages} {086009} (\bibinfo {year} {2005})},\ \Eprint
  {http://arxiv.org/abs/hep-th/0506184} {arXiv:hep-th/0506184} \BibitemShut
  {NoStop}%
\bibitem [{\citenamefont {Mas}\ and\ \citenamefont
  {Tarrio}(2007)}]{Mas:2007ng}%
  \BibitemOpen
  \bibfield  {author} {\bibinfo {author} {\bibfnamefont {J.}~\bibnamefont
  {Mas}}\ and\ \bibinfo {author} {\bibfnamefont {J.}~\bibnamefont {Tarrio}},\
  }\href@noop {} {\bibfield  {journal} {\bibinfo  {journal} {JHEP}\ }\textbf
  {\bibinfo {volume} {05}},\ \bibinfo {pages} {036} (\bibinfo {year} {2007})},\
  \Eprint {http://arxiv.org/abs/hep-th/0703093} {arXiv:hep-th/0703093}
  \BibitemShut {NoStop}%
\bibitem [{\citenamefont {Forster}(1975)}]{Forster:1975}%
  \BibitemOpen
  \bibfield  {author} {\bibinfo {author} {\bibfnamefont {D.}~\bibnamefont
  {Forster}},\ }\href@noop {} {\emph {\bibinfo {title} {{Hydrodynamic
  fluctuations, broken symmetry, and correlation functions}.}}}\ (\bibinfo
  {publisher} {W. A. Benjamin, Inc.: Reading, Mass., Frontiers in Physics.
  Volume 47},\ \bibinfo {year} {1975})\BibitemShut {NoStop}%
\bibitem [{\citenamefont {Yaffe}(1992)}]{Yaffe:1992}%
  \BibitemOpen
  \bibfield  {author} {\bibinfo {author} {\bibfnamefont {L.~G.}\ \bibnamefont
  {Yaffe}},\ }\href@noop {} {\emph {\bibinfo {title} {{Hydrodynamic
  Fluctuations in Relativistic Theories}}}}\ (\bibinfo  {publisher}
  {unpublished},\ \bibinfo {year} {1992})\BibitemShut {NoStop}%
\bibitem [{\citenamefont {Son}\ and\ \citenamefont
  {Starinets}(2002)}]{Son:2002sd}%
  \BibitemOpen
  \bibfield  {author} {\bibinfo {author} {\bibfnamefont {D.~T.}\ \bibnamefont
  {Son}}\ and\ \bibinfo {author} {\bibfnamefont {A.~O.}\ \bibnamefont
  {Starinets}},\ }\href@noop {} {\bibfield  {journal} {\bibinfo  {journal}
  {JHEP}\ }\textbf {\bibinfo {volume} {09}},\ \bibinfo {pages} {042} (\bibinfo
  {year} {2002})},\ \Eprint {http://arxiv.org/abs/hep-th/0205051}
  {arXiv:hep-th/0205051} \BibitemShut {NoStop}%
\bibitem [{\citenamefont {Kaminski}\ \emph {et~al.}(2010)\citenamefont
  {Kaminski}, \citenamefont {Landsteiner}, \citenamefont {Mas}, \citenamefont
  {Shock},\ and\ \citenamefont {Tarrio}}]{Kaminski:2009dh}%
  \BibitemOpen
  \bibfield  {author} {\bibinfo {author} {\bibfnamefont {M.}~\bibnamefont
  {Kaminski}}, \bibinfo {author} {\bibfnamefont {K.}~\bibnamefont
  {Landsteiner}}, \bibinfo {author} {\bibfnamefont {J.}~\bibnamefont {Mas}},
  \bibinfo {author} {\bibfnamefont {J.~P.}\ \bibnamefont {Shock}}, \ and\
  \bibinfo {author} {\bibfnamefont {J.}~\bibnamefont {Tarrio}},\ }\href
  {\doibase 10.1007/JHEP02(2010)021} {\bibfield  {journal} {\bibinfo  {journal}
  {JHEP}\ }\textbf {\bibinfo {volume} {02}},\ \bibinfo {pages} {021} (\bibinfo
  {year} {2010})},\ \Eprint {http://arxiv.org/abs/0911.3610} {arXiv:0911.3610
  [hep-th]} \BibitemShut {NoStop}%
\bibitem [{\citenamefont {Gursoy}(2010)}]{Gursoy:2010jh}%
  \BibitemOpen
  \bibfield  {author} {\bibinfo {author} {\bibfnamefont {U.}~\bibnamefont
  {Gursoy}},\ }\href@noop {} {\  (\bibinfo {year} {2010})},\ \Eprint
  {http://arxiv.org/abs/1007.0500} {arXiv:1007.0500 [hep-th]} \BibitemShut
  {NoStop}%
\bibitem [{\citenamefont {{Kapusta, J.I. and Gale, C.}}(2006)}]{Kapusta2006}%
  \BibitemOpen
  \bibfield  {author} {\bibinfo {author} {\bibnamefont {{Kapusta, J.I. and
  Gale, C.}}},\ }\href@noop {} {\emph {\bibinfo {title} {{Finite-Temperature
  Field Theory: Principles and Applications}}}}\ (\bibinfo  {publisher} {{2nd
  edition, Cambridge Monographs on Mathematical Physics. Cambridge University
  Press, Cambridge}},\ \bibinfo {year} {2006})\BibitemShut {NoStop}%
\bibitem [{\citenamefont {Arnold}\ \emph {et~al.}(2006)\citenamefont {Arnold},
  \citenamefont {Dogan},\ and\ \citenamefont {Moore}}]{Arnold:2006fz}%
  \BibitemOpen
  \bibfield  {author} {\bibinfo {author} {\bibfnamefont {P.~B.}\ \bibnamefont
  {Arnold}}, \bibinfo {author} {\bibfnamefont {C.}~\bibnamefont {Dogan}}, \
  and\ \bibinfo {author} {\bibfnamefont {G.~D.}\ \bibnamefont {Moore}},\ }\href
  {\doibase 10.1103/PhysRevD.74.085021} {\bibfield  {journal} {\bibinfo
  {journal} {Phys. Rev.}\ }\textbf {\bibinfo {volume} {D74}},\ \bibinfo {pages}
  {085021} (\bibinfo {year} {2006})},\ \Eprint
  {http://arxiv.org/abs/hep-ph/0608012} {arXiv:hep-ph/0608012} \BibitemShut
  {NoStop}%
\bibitem [{\citenamefont {Kovtun}\ and\ \citenamefont
  {Starinets}(2006)}]{Kovtun:2006pf}%
  \BibitemOpen
  \bibfield  {author} {\bibinfo {author} {\bibfnamefont {P.}~\bibnamefont
  {Kovtun}}\ and\ \bibinfo {author} {\bibfnamefont {A.}~\bibnamefont
  {Starinets}},\ }\href {\doibase 10.1103/PhysRevLett.96.131601} {\bibfield
  {journal} {\bibinfo  {journal} {Phys. Rev. Lett.}\ }\textbf {\bibinfo
  {volume} {96}},\ \bibinfo {pages} {131601} (\bibinfo {year} {2006})},\
  \Eprint {http://arxiv.org/abs/hep-th/0602059} {arXiv:hep-th/0602059}
  \BibitemShut {NoStop}%
\bibitem [{\citenamefont {Buchel}\ \emph {et~al.}(2005)\citenamefont {Buchel},
  \citenamefont {Liu},\ and\ \citenamefont {Starinets}}]{Buchel:2004di}%
  \BibitemOpen
  \bibfield  {author} {\bibinfo {author} {\bibfnamefont {A.}~\bibnamefont
  {Buchel}}, \bibinfo {author} {\bibfnamefont {J.~T.}\ \bibnamefont {Liu}}, \
  and\ \bibinfo {author} {\bibfnamefont {A.~O.}\ \bibnamefont {Starinets}},\
  }\href {\doibase 10.1016/j.nuclphysb.2004.11.055} {\bibfield  {journal}
  {\bibinfo  {journal} {Nucl. Phys.}\ }\textbf {\bibinfo {volume} {B707}},\
  \bibinfo {pages} {56} (\bibinfo {year} {2005})},\ \Eprint
  {http://arxiv.org/abs/hep-th/0406264} {arXiv:hep-th/0406264} \BibitemShut
  {NoStop}%
\end{thebibliography}%
\end{document}